\documentclass[a4paper,11pt]{article}
\usepackage{jheppub} 
\usepackage{lineno}
\usepackage{amsmath}
\usepackage{amsthm}
\usepackage{amssymb}
\usepackage{xcolor}
\usepackage{makecell}
\usepackage{threeparttable}
\usepackage{tabularx}
\newcolumntype{Y}{>{\centering\arraybackslash}X}
\definecolor{lightblue}{rgb}{0.36, 0.51, 0.71}
\definecolor{lightred}{rgb}{0.94, 0.50, 0.50}

\theoremstyle{definition}
\newtheorem{DEF}{Definition}
\newtheorem{prop}{Proposition}
\newtheorem{Exam}{Example}
\newcommand{\ms}{\mathrm{ms}}
\newcommand{\ps}{\mathrm{ps}}

\title{\boldmath Notes on Fourier transform and its application to three-point momentum-space integrals}

\author{Xuhang Jiang}
\affiliation{Institute of Theoretical Physics, Chinese Academy of Sciences, Beijing 100190, China}

\emailAdd{xhjiang@itp.ac.cn}

\abstract{The Fourier transform of two-point momentum-space Feynman integrals with massless propagators and two off-shell legs can be used to prove identities between their periods, exemplified by the glue-and-cut identity. We generalize this framework to massless momentum-space Feynman integrals with three off-shell legs and obtain a similar family of identities that can be used to calculate these integrals, especially for a non-planar subset of them, which naturally arise in the off-shell Sudakov form factors.}

\begin{document}
\maketitle
\flushbottom

\section{Introduction}
\label{sec:intro}
Feynman integrals (FIs), which form the foundation of perturbative calculations in quantum field theory, satisfy a wide variety of internal identities among themselves. Such identities play a crucial rule in the reduction and evaluation of FIs. The most common and universal identities among general FIs are the Integration-By-Part (IBP) relations~\cite{Chetyrkin:1981qh,Tkachov:1981wb}, which are linear relations among FIs and have been explored in great depth by many different approaches in the last few decades~\cite{Laporta:2000dsw,Laporta:2000dc,Larsen:2015ped,Ita:2015tya,Bohm:2017qme,Mastrolia:2018uzb,Frellesvig:2019uqt,Song:2025pwy,Zeng:2025xbh}. Besides the IBP relations, additional nontrivial identities exist, especially for integrals with additional symmetries. One such example is the (generalized) magic identity for conformal integrals~\cite{Drummond:2006rz,Caron-Huot:2021usw,He:2025vqt}, which relate different four-point conformal integrals in momentum space with each other. Many identities between four-point conformal integrals, e.g. twist identities\footnote{The twist identity is an alternative formulation of the magic identity.} and planar duality, have been investigated systematically in the framework of graphical function~\cite{Schnetz:2008mp,Borinsky:2021gkd,Borinsky:2022lds,Hu:2018liw,Schnetz:2025mjw}. They are mostly used to study the periods of Feynman integrals, which are interesting both physically and mathematically~\cite{Broadhurst:1995km,Broadhurst:1996az,Kreimer:1996js,Broadhurst:1996kc,Brown:2009ta,Brown:2010bw}. Another example of amazing identities among conformal integrals is the Basso-Dixon formula for fishnet family, which evaluates to the determinant of ladders~\cite{Basso:2017jwq,Basso:2021omx,Aprile:2023gnh}. These identities are not linear and thus highly non-trivial. The third example is the glue-and-cut identity~\cite{Baikov:2010hf} among p-integrals (massless Feynman integrals with two off-shell legs), which relates different p-integrals through graph operations as the name indicates. By exploiting all these identities, one can not only reduce the integrals to a known basis, but also use them as constraints to obtain analytic results~\cite{Baikov:2010hf,He:2025vqt}. For example, the glue-and-cut identity~\cite{Baikov:2010hf} has been used to calculate p-integrals up to 5 loops~\cite{Georgoudis:2021onj}.

Interestingly, by Fourier transform, the glue-and-cut identity in momentum space can be translated into the periods identity of graphical functions in position space. This can be proved by first Fourier transforming the momentum-space integral with two off-shell legs to position space, completing it to conformal integrals and then using the conformal property to fix any two adjacent vertices to $0$ and $1$. We give an example of this correspondence in Table.~\ref{tab:zigzag}, while referring the proof by Fourier transform to the appendix of \cite{He:2025lzd}.
\begin{table}[htbp]
        \centering
        \begin{threeparttable}
        \begin{tabularx}{\textwidth}{Y|Y}
            \hline\hline
             position space & loop momentum space \\
             \hline
            \thead{$\vcenter{\hbox{\includegraphics[scale=0.45]{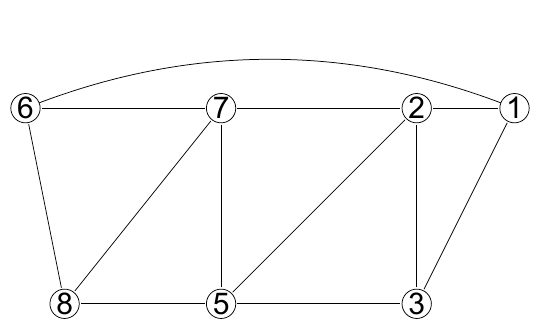}}}$ } & \thead{$\vcenter{\hbox{\includegraphics[scale=0.45]{fig/d2int1fig.pdf}}}$ } \\
            \hline
            \thead{fix $x_6=0$ and $x_1=1$} & \thead{cut line 61 and assign it an \\ off-shell momentum\tnote{1}  $\,\,q^2$} \\
            \hline
            \thead{$x_4$ has been fixed to infinity and the edges joining\\ degree-3 vertices and $x_4$ have been removed} & \thead{the external momenta attached to \\ degree-3 vertices have been set to 0}\\
            \hline
            result: $168\zeta_{9}$ & result: $168\zeta_{9}$ \\
            \hline\hline
        \end{tabularx}
        \begin{tablenotes}
            \item[1] Note that in Euclidean space $q^2=1$, while in Minkowski space, $q^2=-1$.
        \end{tablenotes}
        \caption{The same diagram has different meanings in momentum space and position space. They are related by Fourier transform. Note that we can set any two vertices of an edge to $0$ and $1$ in position space, which corresponds to cutting the edge and assigning an off-shell momentum to it in momentum space.}\label{tab:zigzag}
        \end{threeparttable}
\end{table}

In this work, we extend this correspondence to integrals with three off-shell legs and refer to it as the generalized Fourier identity, which is distinguished from the Fourier identity of periods introduced in \cite{Schnetz:2008mp,Broadhurst:1986bx,Broadhurst:1995km}. The generalized Fourier identities are also formulated by a set of graph rules. They relate Feynman integrals with massless propagators and three off-shell legs in momentum space to their counterparts in position space, which are conformal integrals with four external vertices fixed. The graphs in momentum space and position space follow different Feynman rules as listed in Table~\ref{tab:feynrule}.
\begin{table}[htbp]
    \centering
    \begin{tabular}{ccc}
    \hline\hline
         & Feynman rule & integration measure \\
    \hline
      position space   & $\vcenter{\hbox{\includegraphics{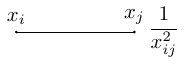}}}$ & $\int\frac{\mathrm{d}^{4}x}{\pi^2}$ \\
      momentum space  & $\vcenter{\hbox{\includegraphics{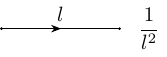}}}$ & $\int\frac{\mathrm{d}^{4}l}{\pi^2}$\\
    \hline\hline
    \end{tabular}
    \caption{The Feynman rules and integration measures in position space and momentum space.}
    \label{tab:feynrule}
\end{table}
The normalization factor $1/\pi^2$ in the integration measure is for the convenience of Fourier transform, which is also a common convention in the literature. The propagators in these two spaces are related by the general Fourier transform in $d=4-2\epsilon$ dimension
\begin{equation}\label{eq:fourier}
    \frac{1}{(k^2)^{\alpha}}=\frac{\Gamma(2-\alpha-\epsilon)}{\Gamma(\alpha)}\int\frac{\mathrm{d}^{d}x}{\pi^{d/2}}\frac{e^{2ik\cdot x}}{(x^2)^{2-\alpha-\epsilon}},
\end{equation}
which simplifies to
\begin{equation}\label{eq:fourier4d}
    \frac{1}{k^2}=\int\frac{\mathrm{d}^{4}x}{\pi^{2}}\frac{e^{2ik\cdot x}}{x^2}
\end{equation}
in four dimension for $\alpha=1$. The generalized Fourier identity is closely related to the planar duality~\cite{Golz:2015rea,Borinsky:2021gkd,Borinsky:2022lds}, which relates a graphical function to the one associated with its dual graph. These two concepts coincide in many cases but different in general.
While the main result of this paper might be already known to some experts, we are not aware of any previous publication where it has been explicitly formulated. 

This paper is organized as follows. In section~\ref{sec:identity}, we introduce the generalized Fourier identities, which relate momentum-space integrals to their position-space counterparts. They can be systematically generated by graph rules. We give a proof of these identities by Fourier transform, Mellin transform and star-triangle relations~\cite{Vasiliev:1981dg,Kazakov:1983dyk,Isaev:2007uy,Chicherin:2012yn}. We will mainly apply it to a particular family of non-planar momentum-space integrals, which naturally arise in the off-shell Sudakov form factors~\cite{Belitsky:2022itf,Belitsky:2023ssv}. In section~\ref{app:proofbyfp}, we discuss the relations between these identities and the planar duality of graphical functions. In section~\ref{sec:conclusion}, we summarize and conclude.

\section{Generalized Fourier identity and its application}\label{sec:identity}

In this section, we discuss the main result of this article. It states a kind of identities between momentum-space integrals with massless propagators and three off-shell legs with graphical functions~\cite{Schnetz:2013hqa,Borinsky:2021gkd,Schnetz:2021ebf}, which also arise as four-point conformal integrals. We call such identities the generalized Fourier identities to distinguish it from the Fourier identities for periods of Feynman integrals~\cite{Schnetz:2008mp}. Then we apply them to a set of non-planar integrals appearing in the off-shell Sudakov form factors.

\subsection{Generalized Fourier identity}

\begin{figure}[htbp]
\centering
\includegraphics[width=.32\textwidth]{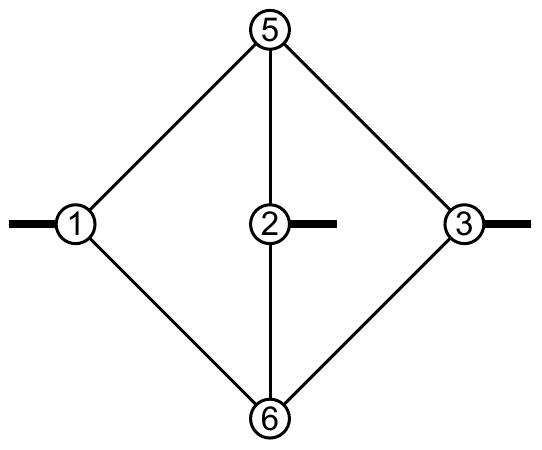}
\qquad
\includegraphics[width=.27\textwidth]{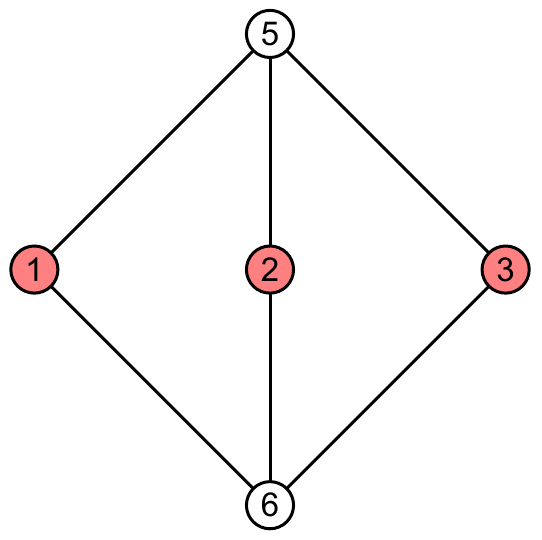}
\caption{Two examples of integrals in different representations. The left diagram is in momentum-space with three external legs, while the right diagram is in position space with three external vertices colored pink.\label{fig:example}}
\end{figure}

We begin by setting up our notations here. The propagators of all the integrals considered below are massless and depicted as thin lines. To distinguish momentum-space integrals from position-space integrals, the former will be dressed up with thick lines indicating off-shell external momenta that always flow outward. Thus, a three-point momentum-space integral will carry three thick lines attached to distinct vertices. For position space integrals, the external points (positions that are fixed) are drawn as pink vertices. All vertices in both the momentum- and position-space integrals are labeled by numbers. Momentum-space integrals are integrated over independent loop momenta flowing along {\it edges}, while position-space integrals are integrated over all internal {\it vertices} (which are not colored). An illustrative example of the notation is provided in Figure~\ref{fig:example}. Although we will show later that the two integrals in Figure~\ref{fig:example} equal to each other after identification of kinematics, they are actually very different according to the defined Feynman rules.

The identities we will discuss are most easily formulated by graph rule.  The graphs we exploit\footnote{These graphs are inspired and closely related to a set of general $f$-integrands or $f$-graphs used to bootstrap the four-point correlators in $\mathcal{N}=4$ super-Yang-Mills theory~\cite{Eden:2011we,Eden:2012tu,Bourjaily:2011hi,Bourjaily:2015bpz,Bourjaily:2016evz,He:2024cej,Bourjaily:2025iad}. However, our discussion is independent of the knowledge of $f$-graphs.} satisfy the following properties:
\begin{enumerate}
    \item The edges can be solid or dashed. An edge must be attached to two distinct vertices (no tadpole). Two solid edges can not be attached to the same two vertices (no bubble for solid lines).
    \item The degree of a vertex is the number of solid lines minus the number of dashed lines attached to the vertex. All vertices in the graph have degree 4.
    \item The dashed lines of the graph (if there are any dashed lines in it) can all be removed by removing a single vertex of the graph.
\end{enumerate}
We denote the set of such graphs as $\mathcal{G}$. Graphs satisfying the first two conditions naturally encode the integrands of conformal integrals in position space where one integrates over all vertices. Solid lines denote propagator denominators, while dashed lines represent the numerators. The degree-4 vertex condition ensures the conformal invariance. The integral is well-defined after fixing the conformal invariance by sending three vertices to $0,1$ and $\infty$. The vertex mapped to $\infty$ and all edges attached it can be removed. Then by the third condition above, one can always choose some vertex to be placed at $\infty$ and the remaining graph is free of dashed lines. This is a necessary condition for the identities to hold. Then we are ready to present the graph rule as follows
\begin{DEF}\label{def:Gp}
    Considering a graph $G\in\mathcal{G}$, choose a vertex in $G$ and send it to $\infty$ so that after removing this vertex and edges attached to it, the reduced graph $G^{\prime}$ is free of dashed lines.
\end{DEF}
\begin{DEF}\label{def:Gmsps}
    Cut two consecutive edges $[v_{1},v_{2}]$ and $[v_{2},v_{3}]$ of $G^{\prime}$ and give the edge $[v_{1},v_{2}]$ an external momentum flow $p_{1}$, the edge $[v_{2},v_{3}]$ another momentum flow $p_{2}$. Take this cut graph as {\it momentum-space} integral and denote it as $G^{\prime}_{\ms}(\frac{p_{1}^{2}}{q^{2}},\frac{p_{2}^{2}}{q^{2}})$ with $q=-p_{1}-p_{2}$. In the same time, consider original $G^{\prime}$, fix the coordinate of above three vertex\footnote{$x_{i}$ is defined as the coordinate of $v_{i}$.} $x_{1}=1$, $x_{2}=z$, $x_{3}=0$ and remove edges $[v_{1},v_{2}]$, $[v_{2},v_{3}]$. Take this reduced graph as {\it position-space} integral and denote it as $G^{\prime}_{\ps}(z,\bar{z})$.
\end{DEF}
\begin{prop}\label{prop:id}
    If $G^{\prime}_{\ms}(\frac{p_{1}^{2}}{q^{2}},\frac{p_{2}^{2}}{q^{2}})$ is finite for generic kinematics $p_{1}^{2},p_{2}^{2},q^{2}\ne 0$ (no sub-divergence), then $G^{\prime}_{\ms}(\frac{p_{1}^{2}}{q^{2}},\frac{p_{2}^{2}}{q^{2}})=\frac{1}{(q^{2})^{2}}G^{\prime}_{\ps}(z,\bar{z})$ where $\frac{p_{1}^{2}}{q^{2}}=z\bar{z}, \frac{p_{2}^{2}}{q^{2}}=(1-z)(1-\bar{z})$.
\end{prop}
In position space, we denote the distance between two vertices $v_{i}$ and $v_{j}$ as $x_{ij}^{2}=(x_{i}-x_{j})^2$. Fixing $x_{1}=1$, $x_{2}=z$ and $x_{3}=0$ represents $x_{12}^{2}=(1-z)(1-\bar{z})$ and $x_{23}^{2}=z\bar{z}$. Note that $p_{1}^{2}/p_{2}^{2}=x_{23}^{2}/x_{12}^{2}$, which makes explicit the duality between momentum space and position space. It also indicates a proof by Fourier transform. Furthermore, this result can be generalized to $d$ dimension while keeping the conformal property of the integral by deforming the edge weights\footnote{The author thanks Oliver Schnetz for pointing this out. The edge weights are the exponents of denominators of propagators.}. In this case, an additional constraint must be imposed on the edge weights of the two edges being cut, as we will discuss later. Before proving this proposition at section~\ref{sec:proof}, we present some explicit examples.

\begin{Exam}\label{ex:K5}
Let us start from the five-point complete graph $G$. The reduced graph of $G$ is $G^{\prime}$ by sending $v_{4}$ (colored orange) to $\infty$,
\begin{equation}\label{eq:K5}
       G= \vcenter{\hbox{\includegraphics[scale=0.4]{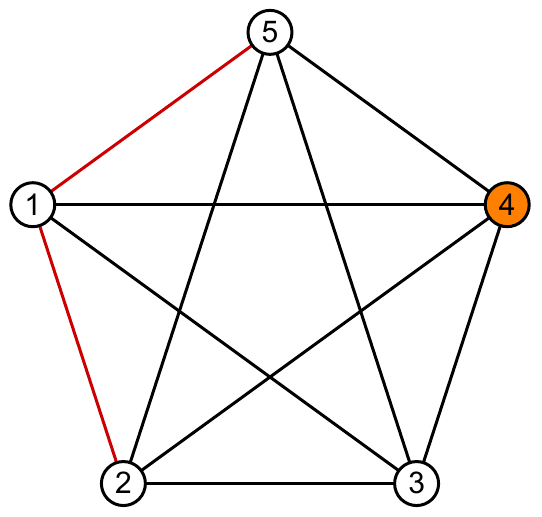}}},  \quad G^{\prime}= \vcenter{\hbox{\includegraphics[scale=0.33]{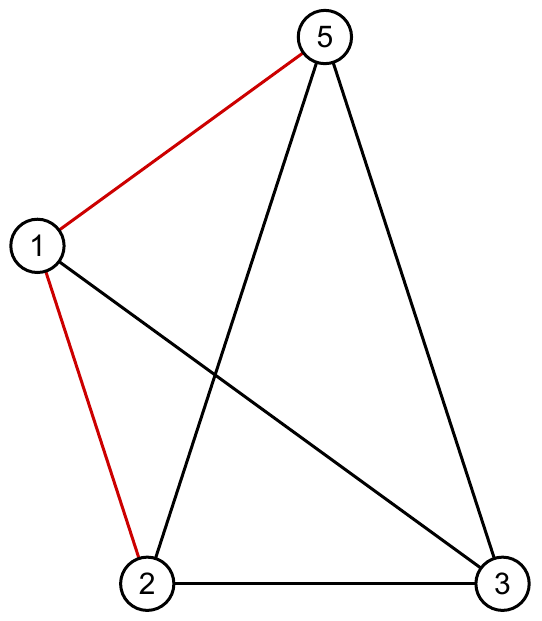}}}.
\end{equation}
Then $G^{\prime}_{\ms}(\frac{p_{1}^{2}}{q^{2}},\frac{p_{2}^{2}}{q^{2}})$ is
\begin{equation}
    \vcenter{\hbox{\includegraphics[scale=0.3]{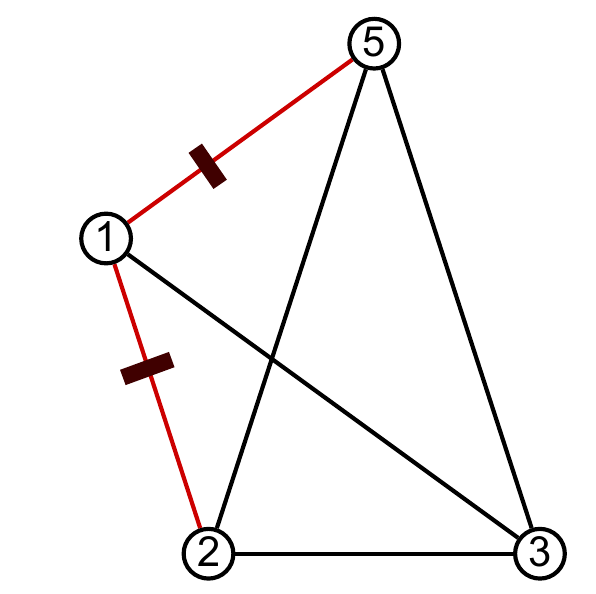}}}\Rightarrow G^{\prime}_{\ms}(\frac{p_{1}^{2}}{q^{2}},\frac{p_{2}^{2}}{q^{2}})=\vcenter{\hbox{\includegraphics[scale=0.4]{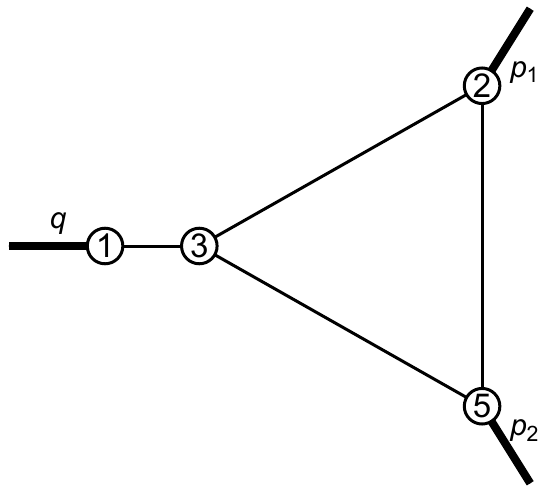}}}.
\end{equation}
Its result is well-known~\cite{Usyukina:1992jd,Loebbert:2019vcj} to be
\begin{equation}
    G^{\prime}_{\ms}(\frac{p_{1}^{2}}{q^{2}},\frac{p_{2}^{2}}{q^{2}})=\frac{1}{(q^2)^2}\frac{4D_{2}(z,\bar{z})}{z-\bar{z}},
\end{equation}
where $D_{2}(z,\bar{z})$ is the famous Bloch-Wigner function whose expression is
\begin{equation}
    D_2(z,\bar{z})=\log|z|\mathrm{Im}\log(1-z) + \mathrm{Im}\mathrm{Li}_{2}(z).
\end{equation}
Now we consider the position space 
\begin{equation}\label{eq:oneloopbox}
    \vcenter{\hbox{\includegraphics[scale=0.3]{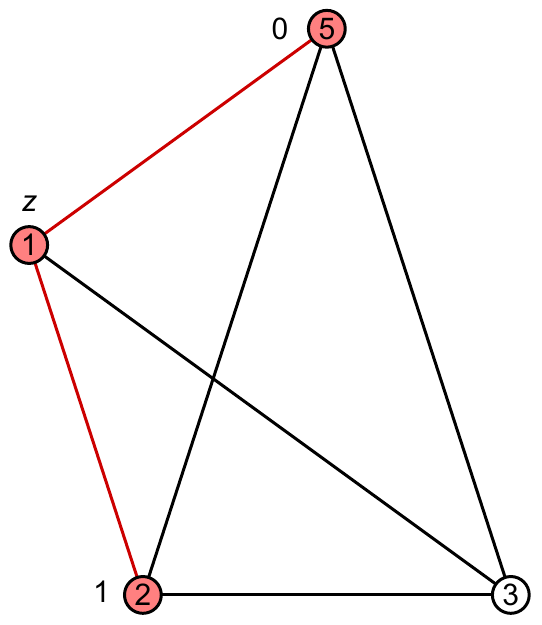}}}\Rightarrow G^{\prime}_{\ps}(z,\bar{z})=\vcenter{\hbox{\includegraphics[scale=0.4]{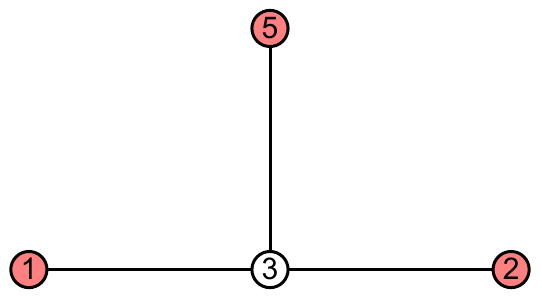}}}=\frac{4D_{2}(z,\bar{z})}{z-\bar{z}},
\end{equation}
where the edge $[v_{2},v_{5}]$ is also removed since the distance $x_{25}^{2}=1$. It is the same with one-loop triangle momentum-space integral and can be computed as graphical function~\cite{Schnetz:2013hqa}. It is then obvious that $G^{\prime}_{\ms}(\frac{p_{1}^{2}}{q^{2}},\frac{p_{2}^{2}}{q^{2}})=\frac{1}{(q^{2})^{2}}G^{\prime}_{\ps}(z,\bar{z})$ in the example above. Due to the symmetry of $G^{\prime}$, taking any two consecutive edges will arrive at the same result.
\end{Exam}

\begin{Exam}\label{ex:threeloop}
    We give another example to show that the generalized Fourier identity is related to the planar duality when the planar dual of the momentum-space integral exists. Consider the following graph $G$,
    \begin{equation}
        G= \vcenter{\hbox{\includegraphics[scale=0.4]{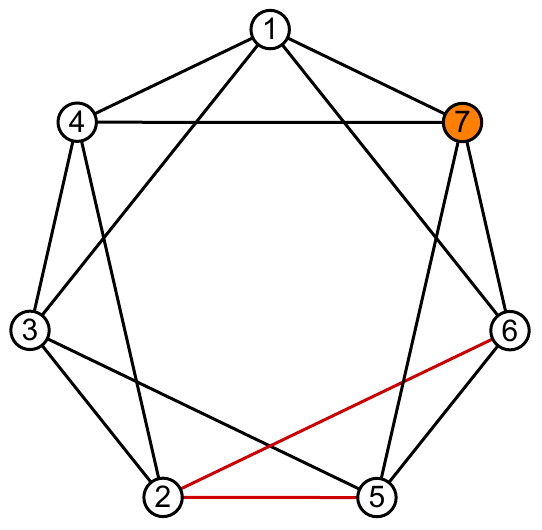}}}.
    \end{equation}
    We will send $x_{7}$ (colored orange) to $\infty$ and cut edges $[v_{5},v_{2}]$ and $[v_{2},v_{6}]$. The resulting integrals in momentum space and position space are
    \begin{equation}
        G^{\prime}_{\ms}(\frac{p_{1}^{2}}{q^2},\frac{p_{2}^{2}}{q^{2}})=\vcenter{\hbox{\includegraphics[scale=0.4]{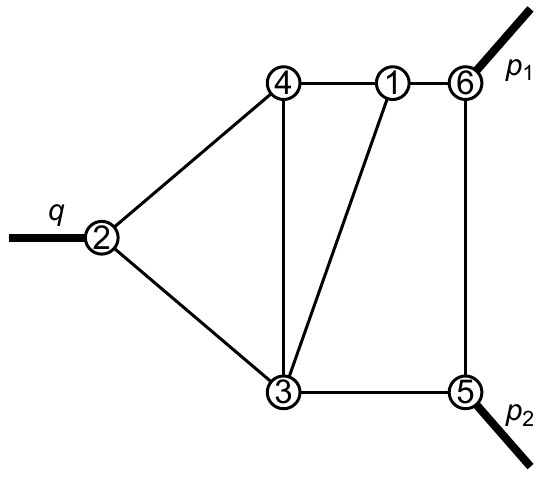}}}, \quad G^{\prime}_{\ps}(z,\bar{z})=\vcenter{\hbox{\includegraphics[scale=0.4]{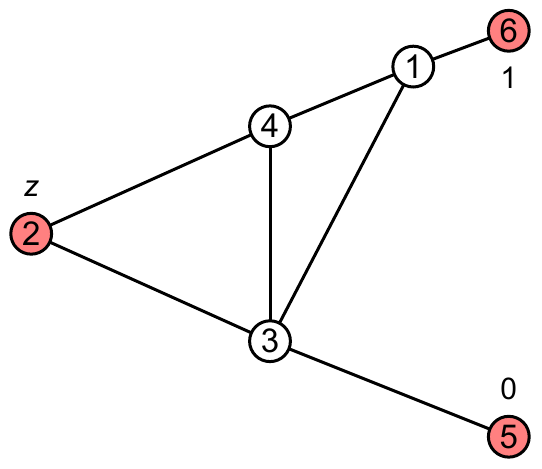}}}
    \end{equation}
    where in the right graph we have omitted the edge between $v_{5}$ ($0$) and $v_{6}$ ($1$). $G^{\prime}_{\ms}(\frac{p_{1}^{2}}{q^2},\frac{p_{2}^{2}}{q^{2}})$ can be evaluated by the package \texttt{HyperInt}~\cite{Panzer:2014caa} and $G^{\prime}_{\ps}(z,\bar{z})$ can be evaluated by another package \texttt{HyperlogProcedures}~\cite{hyperlogprocedures}. The two results agree with each other and satisfy the proposition. Note that $G^{\prime}_{\ps}$ is {\it not} the dual graph of $G^{\prime}_{\ms}$. Since $G^{\prime}_{\ms}$ can also be represented by its dual graph, setting $q^2=-1$, we derive the following identity between two integrals in position space, 
    \begin{equation}
        \vcenter{\hbox{\includegraphics[scale=0.35]{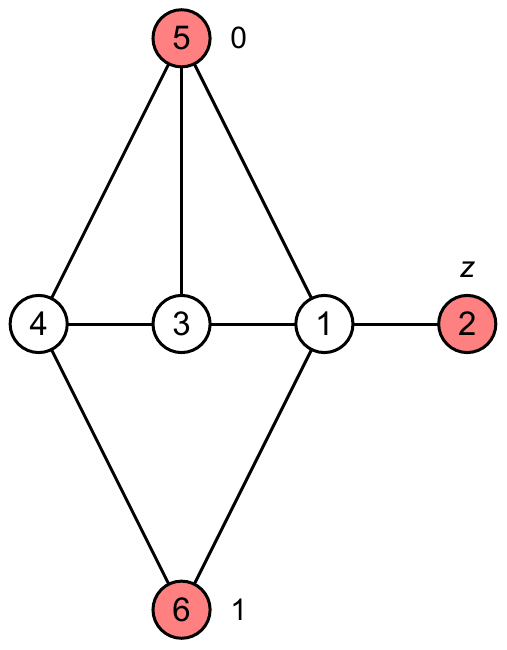}}} \quad =\quad \vcenter{\hbox{\includegraphics[scale=0.35]{fig/threeloop_zigzag_ps.pdf}}}
    \end{equation}
    where the left diagram is the dual representation of $G^{\prime}_{\ms}(\frac{p_{1}^{2}}{q^2},\frac{p_{2}^{2}}{q^{2}})$.
    This is exactly the planar duality relation for graphical functions\footnote{The left graph is exactly dual to the right after recovering the edge between vertex $v_{5}$ ($0$) with $v_{6}$ ($1$) in the right graph.}~\cite{Golz:2015rea,Borinsky:2021gkd,Borinsky:2022lds}. We will discuss more in section~\ref{app:proofbyfp} about the relations between the generalized Fourier identity and planar duality which is proved in Feynman parameterization~\cite{Golz:2015rea}.
\end{Exam}

\begin{Exam}
    Let us still take the diagram in eq.~\eqref{eq:K5} as an example, but with the edge weights being deformed from $1$. That is, the Feynman rule for propagators will be $1/(l^2)^{\nu}$ with $\nu$ being general complex parameter instead of 1. The edge weights are different in momentum space (depicted in color {\color{lightred} pink}) and in position space (depicted in color {\color{lightblue} blue}) as the general Fourier transform eq.~\eqref{eq:fourier} indicates. They have been marked in different color in the following diagram
    \begin{equation}\label{eq:K5w}
        G=\vcenter{\hbox{\includegraphics[scale=0.4]{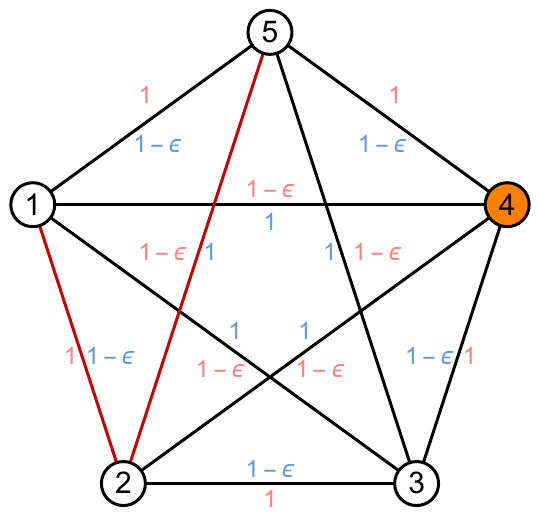}}}.
    \end{equation}
    We denote the weights of the two edges being cut ($[v_1,v_2], [v_2,v_5]$ colored red in eq.~\eqref{eq:K5w}) in momentum space as $\nu_1^{\prime}$ and $\nu_2^{\prime}$. $\nu_1^{\prime}+\nu_2^{\prime}=d/2=2-\epsilon$ is a necessary condition for the generalized identity to hold. The corresponding edge weights in position space are $\nu_{i}=d/2-\nu_{i}^{\prime}$. So it follows that this condition holds in position space as well. This is why we cut $[v_1,v_2], [v_2,v_5]$ instead of $[v_2,v_1], [v_1,v_5]$. It can also be seen that, the total edge weight of the edges attached to each vertex equals to $d=4-2\epsilon$, which is another necessary condition which guarantees the conformal property of the integral. Now this identity can be stated as
    \begin{equation}
        G^{\prime}_{\ms}(\frac{p_{1}^{2}}{q^{2}},\frac{p_{2}^{2}}{q^{2}})=\frac{1}{(q^{2})^{d/2}}G^{\prime}_{\ps}(z,\bar{z})
    \end{equation}
    Let us check it by explicit evaluation in momentum space and position space. First, in momentum space,
    \begin{equation}
        G^{\prime}_{\ms}(\frac{p_{1}^{2}}{q^{2}},\frac{p_{2}^{2}}{q^{2}})=\vcenter{\hbox{\includegraphics[scale=0.4]{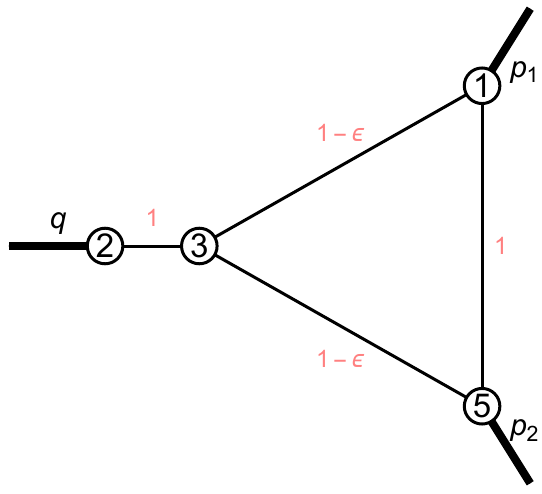}}}.
    \end{equation}
    It can be evaluated in Feynman parameterization as
    \begin{equation}
        G^{\prime}_{\ms}(\frac{p_{1}^{2}}{q^{2}},\frac{p_{2}^{2}}{q^{2}})=\frac{1}{(q^{2})^{2-\epsilon}}\frac{1}{\Gamma(1-\epsilon)}\int[\mathrm{d}\alpha_{1}\mathrm{d}\alpha_{2}\mathrm{d}\alpha_{3}]\frac{\alpha_{2}^{-\epsilon}\alpha_{3}^{-\epsilon}}{(\alpha_1\!+\!\alpha_2\!+\!\alpha_3)(\alpha_{2}\alpha_{3}+\frac{p_1^2}{q^2}\alpha_1\alpha_3+\frac{p_2^2}{q^2}\alpha_1\alpha_2)^{1-\epsilon}}
    \end{equation}
    with $z\bar{z}\equiv p_1^2/q^2$, $(1-z)(1-\bar{z})\equiv p_2^2/q^2$. $[\mathrm{d}\alpha_{1}\mathrm{d}\alpha_{2}\mathrm{d}\alpha_{3}]$ indicates the integration with a delta function to fix the scaling. Now we turn to the position space,
    \begin{equation}
        G^{\prime}_{\ps}(z,\bar{z})=\vcenter{\hbox{\includegraphics[scale=0.4]{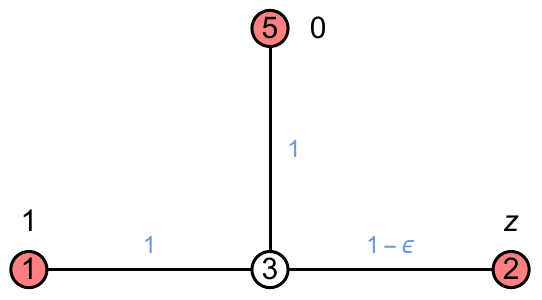}}}.
    \end{equation}
    It can also be evaluated in Feynman parameterization as
    \begin{equation}\label{eq:psweight}
        G^{\prime}_{\ms}(\frac{p_{1}^{2}}{q^{2}},\frac{p_{2}^{2}}{q^{2}})=\frac{1}{\Gamma(1-\epsilon)}\int[\mathrm{d}\alpha_{1}\mathrm{d}\alpha_{2}\mathrm{d}\alpha_{3}]\frac{\alpha_{1}^{-\epsilon}}{(\alpha_1\!+\!\alpha_2\!+\!\alpha_3)^{1-\epsilon}(\alpha_{2}\alpha_{3}+\frac{p_1^2}{q^2}\alpha_1\alpha_3+\frac{p_2^2}{q^2}\alpha_1\alpha_2)}
    \end{equation}
    Performing a Cremona transformation $\alpha_{i}\to 1/\alpha_{i}$ to eq.~\eqref{eq:psweight} and then rescaling $\alpha_{2}\to \alpha_{2}q^2/p_1^2$, $\alpha_{3}\to \alpha_{3}q^2/p_2^2$, it can be shown straightforwardly that 
    \begin{equation}
        G^{\prime}_{\ms}(\frac{p_{1}^{2}}{q^{2}},\frac{p_{2}^{2}}{q^{2}})=\frac{1}{(q^{2})^{2-\epsilon}}G^{\prime}_{\ms}(\frac{p_{1}^{2}}{q^{2}},\frac{p_{2}^{2}}{q^{2}}).
    \end{equation}
    we will prove the general case in section~\ref{sec:proof}.
\end{Exam}

\subsection{proof of the generalized Fourier identities}\label{sec:proof}
\begin{proof}
Now we prove Proposition~\ref{prop:id} by Fourier transform, Mellin transform and star-triangle relation (or called the method of uniqueness)~\cite{Vasiliev:1981dg,Kazakov:1983dyk,Isaev:2007uy,Chicherin:2012yn}. Readers mainly interested in the applications can skip this section. 

We first keep ourselves in four dimensions and the Feynman rules are given in table~\ref{tab:feynrule}. Starting from $G^{\prime}_{\ms}$ with edge $[v_{1},v_{2}]$ and $[v_{2},v_{3}]$ cut, we perform the Fourier transform to every propagator of the momentum-space integral by eq.~\eqref{eq:fourier4d}. Integrating out all the loop momenta will result in a set of delta functions for coordinates by
\begin{equation}
        \int\mathrm{d}^{4}k e^{2ik\cdot x}=\pi^{4}\delta^{(4)}(x).
\end{equation}
These delta functions will be satisfied if we identify the dual coordinates of loop momentum $l$ flowing along edge $[v_{i},v_{j}]$ with $x_{i}-x_{j}$. Thus the resulting position-space integral share the same graph with original momentum-space integral, except that the integrand related to the three vertices $v_{1},v_{2},v_{3}$ where the three external legs are attached is modified. Due to the translation invariance of integral in position space, we fix $x_{3}$ to 0. The resulting integral is then\footnote{The transformation between momentum space and position space can be derived straightforwardly. We refer to \cite{Chetyrkin:1980pr} for some pedagogical examples.}
\begin{equation}\label{eq:fourierlast}
    G^{\prime}_{\ms}(\frac{p_{1}^{2}}{q^2},\frac{p_{2}^{2}}{q^{2}})=\int\frac{\mathrm{d}^{4}x_{1}}{\pi^{2}}\int\frac{\mathrm{d}^{4}x_{2}}{\pi^{2}}e^{2ip_{1}\cdot x_{1}}e^{2iq\cdot x_{2}}\frac{1}{(x_{1}^{2})^{2}}I_{G^{\prime}_{\ps}}(\frac{x_{12}^{2}}{x_{1}^{2}},\frac{x_{2}^{2}}{x_{1}^{2}}).
\end{equation}
$I_{G^{\prime}_{\ps}}$ is the integral of $G^{\prime}_{\ps}$ but with three ``external'' vertices $x_{1},x_{2},x_{3}$ where $x_{3}$ has been fixed to 0. Note that $x_{1}$ and $x_{2}$ need to be further integrated out, so they are not the genuine external points. $I_{G^{\prime}_{\ps}}$ depends only on two cross ratios $\frac{x_{12}^{2}x_{34}^{2}}{x_{13}^{2}x_{24}^{2}}=\frac{x_{12}^{2}}{x_{1}^{2}},\frac{x_{14}^{2}x_{23}^{2}}{x_{13}^{2}x_{24}^{2}}=\frac{x_{2}^{2}}{x_{1}^{2}}$ after we extract a normalization factor $1/(x_{1}^{2})^2$ out\footnote{The power of $x_{1}^{2}$ is fixed by the graph condition. Supposing such diagram has $V$ vertices, then the integration measure has mass dimension $-4V$ due to the vertex degree-4 condition. The number of edges are $4V/2=2V$. However, we have removed two edges as in Definition~\ref{def:Gmsps}, thus the overall mass dimension of the integrand in position space is $-4$, which indicates a normalization factor $1/(x^{2})^2$.}, because the diagram has conformal property and we can always complete this graph to conformal integrals by adding the $x_{4}=\infty$ back. Now we rewrite $I_{G^{\prime}_{\ps}}$ by its Mellin transform:
\begin{equation}\label{eq:mellin}
    I_{G^{\prime}_{\ps}}(\frac{x_{12}^{2}}{x_{1}^{2}},\frac{x_{2}^{2}}{x_{1}^{2}})=\int\frac{\mathrm{d}s}{2\pi i}\int\frac{\mathrm{d}t}{2\pi i}\left(\frac{x_{12}^{2}}{x_{1}^{2}}\right)^{-s}\left(\frac{x_{2}^{2}}{x_{1}^{2}}\right)^{-t}\tilde{I}_{G^{\prime}_{\ps}}(s,t),
\end{equation}
where $\tilde{I}_{G^{\prime}_{\ps}}(s,t)$ is the Mellin transform of $I_{G^{\prime}_{\ps}}(\frac{x_{12}^{2}}{x_{1}^{2}},\frac{x_{2}^{2}}{x_{1}^{2}})$. In eq.~\eqref{eq:fourierlast} and eq.~\eqref{eq:mellin}, we assume that $G^{\prime}_{\ms}(\frac{p_{1}^{2}}{q^2},\frac{p_{2}^{2}}{q^{2}})$ (and thus $I_{G^{\prime}_{\ps}}(\frac{x_{12}^{2}}{x_{1}^{2}},\frac{x_{2}^{2}}{x_{1}^{2}})$ which is the Fourier transform of $G^{\prime}_{\ms}(\frac{p_{1}^{2}}{q^2},\frac{p_{2}^{2}}{q^{2}})$) is finite for general kinematics so that its Mellin transform $\tilde{I}_{G^{\prime}_{\ps}}(s,t)$ and the multiple integration is well-defined. Then we can exchange the integration order,
\begin{equation}
    \begin{aligned}
        G^{\prime}_{\ms}(\frac{p_{1}^{2}}{q^2},\frac{p_{2}^{2}}{q^{2}})&=\!\!\!\int\frac{\mathrm{d}s}{2\pi i}\int\frac{\mathrm{d}t}{2\pi i}\tilde{I}_{G^{\prime}_{\ps}}(s,t)\int\frac{\mathrm{d}^{4}x_{1}}{\pi^{2}}\int\frac{\mathrm{d}^{4}x_{2}}{\pi^{2}}e^{2ip_{1}\cdot x_{1}}e^{2iq\cdot x_{2}}\frac{1}{(x_{1}^{2})^{2}}\left(\frac{x_{12}^{2}}{x_{1}^{2}}\right)^{-s}\left(\frac{x_{2}^{2}}{x_{1}^{2}}\right)^{-t}.
    \end{aligned}
\end{equation}
Now we transform the power of $x_{1}^{2},x_{2}^{2},x_{12}^{2}$ back to momentum space,
\begin{equation}
    \begin{aligned}
        \frac{1}{(x_{1}^{2})^{2-s-t}}&=\frac{\Gamma(s+t)}{\Gamma(2-s-t)}\int\frac{\mathrm{d}^{4}l_{1}}{\pi^2}\frac{e^{2il_{1}\cdot x_{1}}}{(l_{1}^{2})^{s+t}}, \, 
        \frac{1}{(x_{2}^{2})^{t}}=\frac{\Gamma(2-t)}{\Gamma(t)}\int\frac{\mathrm{d}^{4}l_{2}}{\pi^2}\frac{e^{2il_2\cdot x_{2}}}{(l_{2}^{2})^{2-t}} \\
        \frac{1}{(x_{12}^{2})^{s}}&=\frac{\Gamma(2-s)}{\Gamma(s)}\int\frac{\mathrm{d}^{4}l_{3}}{\pi^2}\frac{e^{2il_3\cdot x_{12}}}{(l_{3}^{2})^{2-s}}.
    \end{aligned}
\end{equation}
Note that $s,t$ are complex variables and the Fourier transforms above are defined by analytically continuing in $\alpha$ while keeping $d=4$ fixed in eq.~\eqref{eq:fourier}. After integrating $x_{1},x_{2}$ out, $l_{1}$ is localized to $-l_{3}-p_{1}$ and $l_{2}$ is localized to $l_{3}-q$ by delta functions. Thus,
\begin{equation}\label{eq:fourierintermediate}
    \begin{aligned}
        G^{\prime}_{\ms}(\frac{p_{1}^{2}}{q^2},\frac{p_{2}^{2}}{q^{2}})&=\!\!\!\int\frac{\mathrm{d}s}{2\pi i}\int\frac{\mathrm{d}t}{2\pi i}\tilde{I}_{G^{\prime}_{\ps}}(s,t)c_{\Gamma}\int\frac{\mathrm{d}^{4}l_{3}}{\pi^{2}}\frac{1}{[(l_3+p_1)^{2}]^{s+t}[(l_3-q)^{2}]^{2-t}[l_3^{2}]^{2-s}}
    \end{aligned}
\end{equation}
where
\begin{equation}
    c_{\Gamma}=\frac{\Gamma(s+t)\Gamma(2-t)\Gamma(2-s)}{\Gamma(s)\Gamma(t)\Gamma(2-s-t)}.
\end{equation}
Recall the star-triangle relations for general conformal scalar integrals~\cite{Loebbert:2019vcj},
\begin{equation}\label{eq:startriangle}
    \int\frac{\mathrm{d}^{D}x_{0}}{\pi^{D/2}}\frac{1}{(x_{01}^{2})^{a}(x_{02}^{2})^{b}(x_{03}^{2})^{c}}=\frac{\Gamma(a^{\prime})\Gamma(b^{\prime})\Gamma(c^{\prime})}{\Gamma(a)\Gamma(b)\Gamma(c)}\frac{1}{(x_{12}^{2})^{c^{\prime}}(x_{23}^{2})^{a^{\prime}}(x_{13}^{2})^{b^{\prime}}}, \quad \text{when } a+b+c=D
\end{equation}
where $a^{\prime}=D/2-a,b^{\prime}=D/2-b,c^{\prime}=D/2-c$. Therefore,
\begin{equation}
    \int\frac{\mathrm{d}^{4}l_{3}}{\pi^{2}}\frac{1}{[(l_3+p_1)^{2}]^{s+t}[(l_3-q)^{2}]^{2-t}[l_3^{2}]^{2-s}}=\frac{1}{c_{\Gamma}}\frac{1}{(q^{2})^{2-s-t}(p_{1}^{2})^{t}(p_{2}^{2})^{s}}
\end{equation}
We finally arrive at
\begin{equation}\label{eq:lastsencondstep}
    \begin{aligned}
        G^{\prime}_{\ms}(\frac{p_{1}^{2}}{q^2},\frac{p_{2}^{2}}{q^{2}})&=\frac{1}{(q^{2})^2}\int\frac{\mathrm{d}s}{2\pi i}\int\frac{\mathrm{d}t}{2\pi i}\tilde{I}_{G^{\prime}_{\ps}}(s,t)\left(\frac{p_{2}^{2}}{q^{2}}\right)^{-s}\left(\frac{p_{1}^{2}}{q^{2}}\right)^{-t}
    \end{aligned}
\end{equation}
Compared with eq.~\eqref{eq:mellin}, it can be recognized that the right-hand side is $1/(q^{2})^2I_{G^{\prime}}(z,\bar{z})$ with $x_{1}=1,x_{2}=z,x_{3}=0$, given $p_{1}^{2}/q^{2}=z\bar{z}$ and $p_{2}^{2}/q^{2}=(1-z)(1-\bar{z})$. Then we arrive at our proposition
\begin{equation}
    G^{\prime}_{\ms}(\frac{p_{1}^{2}}{q^{2}},\frac{p_{2}^{2}}{q^{2}})=\frac{1}{(q^{2})^{2}}G^{\prime}_{\ps}(z,\bar{z}).
\end{equation}

Next we can generalize above result to $d$ dimension while keeping the conformal property of the integrals by deforming the edge weights of propagators. The author thanks Oliver Schnetz for pointing the existence of this generalization out. Supposing the edge weight for each edge $e_{i}$ is $\nu_{i}$ in {\it position space}, and the set of edges attached to vertex $v_{i}$ is defined as $E_{i}$, we require
\begin{equation}
    \sum_{e_{i}\in E_{j}}\nu_{i}=d=4-2\epsilon, \,\, \forall j.
\end{equation}
for the conformal property to hold. After performing the general Fourier transform eq.~\eqref{eq:fourier} to original momentum-space integral, one can generalize eq.~\eqref{eq:fourierlast} to
\begin{equation}
    G^{\prime}_{\ms}(\frac{p_{1}^{2}}{q^2},\frac{p_{2}^{2}}{q^{2}})=\int\frac{\mathrm{d}^{d}x_{1}}{\pi^{d/2}}\int\frac{\mathrm{d}^{d}x_{2}}{\pi^{d/2}}e^{2ip_{1}\cdot x_{1}}e^{2iq\cdot x_{2}}\frac{1}{(x_{1}^{2})^{d-\nu_1-\nu_2}}I_{G^{\prime}_{\ps}}(\frac{x_{12}^{2}}{x_{1}^{2}},\frac{x_{2}^{2}}{x_{1}^{2}}).
\end{equation}
where $\nu_1$ and $\nu_2$ are the weights of the two cut edges. Then eq.~\eqref{eq:fourierintermediate} becomes
\begin{equation}
    \begin{aligned}
        G^{\prime}_{\ms}(\frac{p_{1}^{2}}{q^2},\frac{p_{2}^{2}}{q^{2}})&\!=\!\!\!\int\!\!\frac{\mathrm{d}s}{2\pi i}\!\!\int\!\!\frac{\mathrm{d}t}{2\pi i}\tilde{I}_{G^{\prime}_{\ps}}(s,t)c^{\prime}_{\Gamma}\!\!\int\!\frac{\mathrm{d}^{d}l_{3}}{\pi^{d/2}}\frac{1}{[(l_3\!+\!p_1)^{2}]^{s+t+\nu_1+\nu_2-2+\epsilon}[(l_3\!-\!q)^{2}]^{2-t-\epsilon}[l_3^{2}]^{2-s-\epsilon}}
    \end{aligned}
\end{equation}
where
\begin{equation}
    c^{\prime}_{\Gamma}=\frac{\Gamma(s+t+\nu_1+\nu_2-2+\epsilon)\Gamma(2-t-\epsilon)\Gamma(2-s-\epsilon)}{\Gamma(s)\Gamma(t)\Gamma(d-\nu_1-\nu_2-s-t)}.
\end{equation}
Subtleties arise when dealing with the integration of $l_3$ with star-triangle relation, which requires $2-\epsilon+\nu_1+\nu_2=d$, that is, $\nu_1+\nu_2=d/2$. In four dimension, $\nu_i=1$ and $d=4$, this requirement is naturally satisfied. We assume $\nu_1+\nu_2=d/2$ for general case, then $l_3$ can still be directly integrated out as before. We arrive at the generalization of eq.~\eqref{eq:lastsencondstep}
\begin{equation}
    \begin{aligned}
        G^{\prime}_{\ms}(\frac{p_{1}^{2}}{q^2},\frac{p_{2}^{2}}{q^{2}})&=\frac{1}{(q^{2})^{d-\nu_1-\nu_2}}\int\frac{\mathrm{d}s}{2\pi i}\int\frac{\mathrm{d}t}{2\pi i}\tilde{I}_{G^{\prime}_{\ps}}(s,t)\left(\frac{p_{2}^{2}}{q^{2}}\right)^{-s}\left(\frac{p_{1}^{2}}{q^{2}}\right)^{-t}
    \end{aligned}
\end{equation}
As a result,
\begin{equation}
    G^{\prime}_{\ms}(\frac{p_{1}^{2}}{q^{2}},\frac{p_{2}^{2}}{q^{2}})=\frac{1}{(q^{2})^{d/2}}G^{\prime}_{\ps}(z,\bar{z}).
\end{equation}
We finish the proof of Proposition~\ref{prop:id} and its generalization.
\end{proof}

A lot of identities can be systematically generated from proposition above, as long as the graph set $\mathcal{G}$ is given. However, in the following we will focus on its application to a particular class of non-planar momentum-space integrals, which has not been discussed in general in the literature before.
\subsection{The application to the off-shell form factor}\label{sec:offshellff}
In this section, we consider a particular family of non-planar integrals relevant to the off-shell form factor~\cite{Belitsky:2022itf,Belitsky:2023ssv}. Their evaluation in Feynman parameterization is usually time-consuming or, in some cases, infeasible\footnote{The calculations in Feynman parameterization are done in \texttt{HyperInt}~\cite{Panzer:2014caa}. Recently, a new package \texttt{HyperForm}~\cite{Kardos:2025klp} has been presented. It is more efficient in non-trivial examples. Nevertheless, for the integration in Feynman parameterization, another difficulty lies in identifying an efficient linear reduction order, particularly in high-loop cases.}. By contrast, their position-space counterparts, obtained via Proposition~\ref{prop:id}, are readily computable or already known. Consequently, Proposition~\ref{prop:id} serves as an effective tool to handle this class of integrals.

This non-planar family starts from two loops. The first example has been presented in Figure~\ref{fig:example}. Starting from the following $G$ and $G^{\prime}$,
\begin{equation}
   G=\vcenter{\hbox{\includegraphics[scale=0.4]{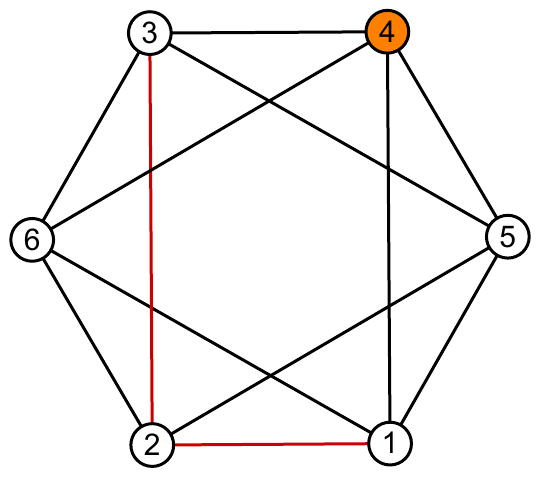}}},\quad G^{\prime}=\vcenter{\hbox{\includegraphics[scale=0.35]{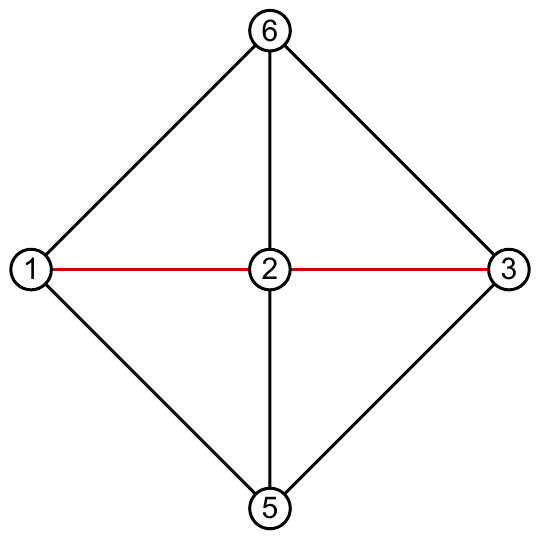}}}
\end{equation}
taking $x_4=\infty$ and cutting two consecutive edges $[v_1,v_2],[v_2,v_3]$, it follows that the integral in momentum space (left in Figure~\ref{fig:example}), denoted as $N^{(2)}_{1}$, equals to the integral in position space (right in Figure~\ref{fig:example}) up to a factor $1/(q^2)^2$, where $x_{1}=1,x_2=z,x_3=0$ and $z,\bar{z}$ are defined by $p_1^2=q^2z\bar{z}$ flowing out of $v_1$ and $p_2^2=q^2(1-z)(1-\bar{z})$ flowing out of $v_{3}$. Furthermore, the integral in position space factorizes into the product of two one-loop conformal box\footnote{In position space, the one-loop conformal box is just the integral presented in eq.~\eqref{eq:oneloopbox}. }. As a result, $N_{1}^{(2)}$ equals to the square of conformal box\footnote{This is also pointed out in \cite{Belitsky:2023ssv}.}, $F_{1}^{2}(z,\bar{z})$, where $F_{L}$ denotes the $L$-loop conformal ladder integral~\cite{Usyukina:1993ch,Broadhurst:1993ib,Broadhurst:2010ds},
\begin{equation}\label{eq:FL}
    F_{L}(z,\bar{z})=\frac{1}{(z-\bar{z})L!}\sum_{j=L}^{2L}\frac{j!}{(j-L)!(2L-j)!}\left(-\log z\bar{z}\right)^{2L-j}\left(\mathrm{Li}_{j}\left(z\right)-\mathrm{Li}_{j}\left(\bar{z}\right)\right).
\end{equation}

\begin{figure}[thbp]
    \centering
    \includegraphics[width=0.27\linewidth]{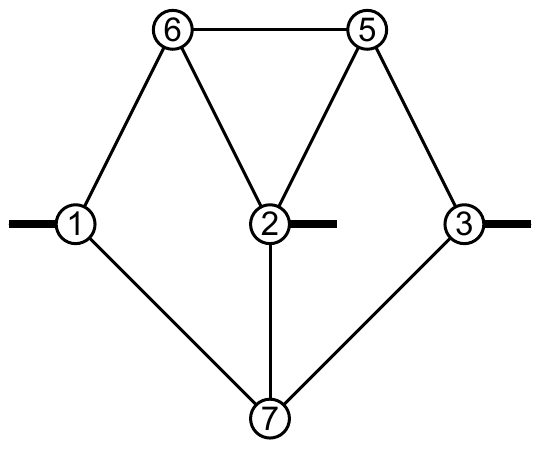}
    \quad
    \includegraphics[width=0.22
    \linewidth]{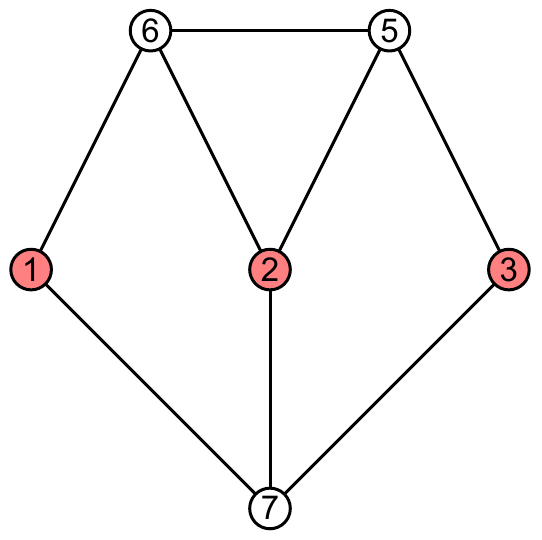}
    \caption{The three-loop non-planar integral which will contribute to the off-shell Sudakov form factor (left) and its counterpart in position space (right).}
    \label{fig:3lnp}
\end{figure}

At three loops, there is also one such integral~\cite{Belitsky:2023ssv}, denoted as $N^{(3)}_{1}$. We depict this integral and its position-space counterpart in Figure~\ref{fig:3lnp}. The kinematic setting is the same as two-loop case. It factorizes into the product of a two-loop conformal ladder and a one-loop conformal box, $F_{1}F_{2}$. 
The graph $G$ which generates these two integrals is depicted in eq.~\eqref{eq:gpentagon}
\begin{equation}\label{eq:gpentagon}
    G=\vcenter{\hbox{\includegraphics[scale=0.4]{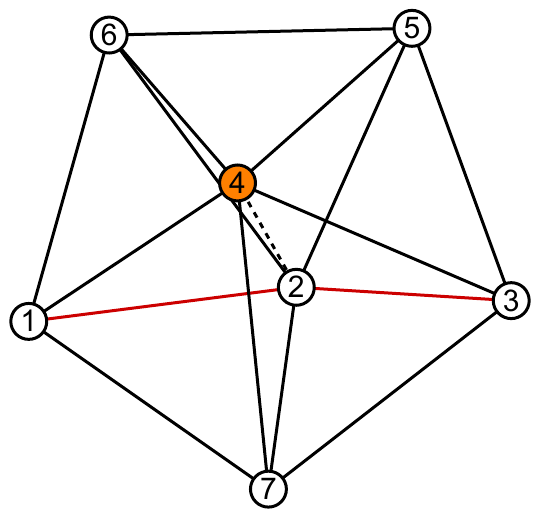}}}
\end{equation}
where the convention is also the same.
Since the conformal ladder family has been calculated to arbitrary loops, we can easily extend above identities between such a non-planar family and product of ladders by generalizing the pentagon in eq.~\eqref{eq:gpentagon} to $n$-gon and cut any two spokes. It can be even further generalized to cases where other finite integrals instead of ladders involve in the factorization, in which case the identities follow in the same way.

At four loops, there are five such integrals in total~\cite{Boels:2017ftb} once the massless external legs are taken off-shell, in direct analog with the procedure employed at three loops~\cite{Belitsky:2023ssv}. We use the same number to denote the off-shell version of the five integrals in \cite{Boels:2017ftb}:
\begin{equation}
    N^{(4)}_{19}\sim I^{(22)}_{19}, \, N^{(4)}_{20}\sim I^{(22)}_{20},\, N^{(4)}_{21}\sim I^{(24)}_{21},\, N^{(4)}_{22}\sim I^{(24)}_{22},\, N^{(4)}_{23}\sim I^{(28)}_{23},
\end{equation}
which are depicted in eq.~\eqref{eq:fourloopnonplanar},
\begin{equation}\label{eq:fourloopnonplanar}
    \begin{aligned}
        N^{(4)}_{19}&=\vcenter{\hbox{\includegraphics[scale=0.3]{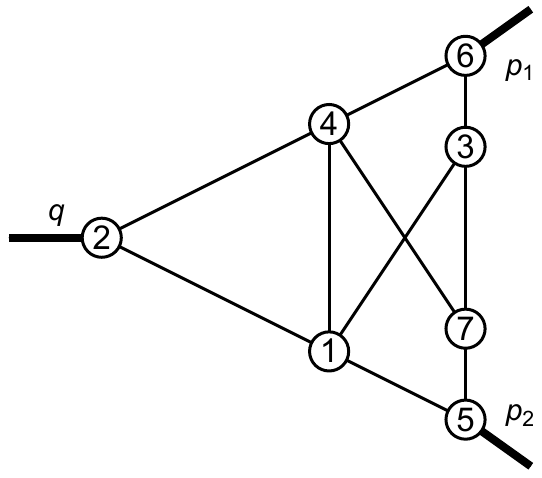}}}, \, N^{(4)}_{20}=\vcenter{\hbox{\includegraphics[scale=0.3]{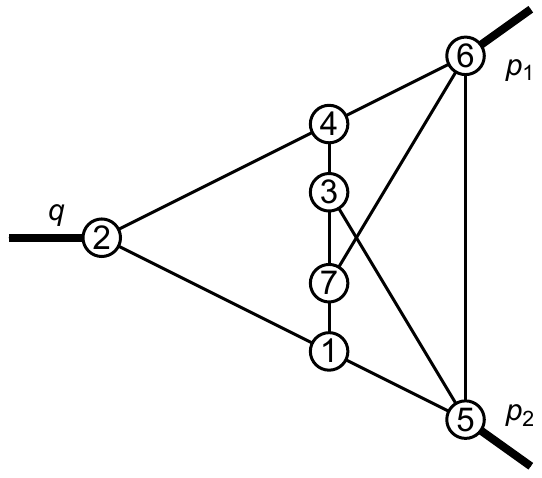}}}, \, 
        N^{(4)}_{21}=\vcenter{\hbox{\includegraphics[scale=0.3]{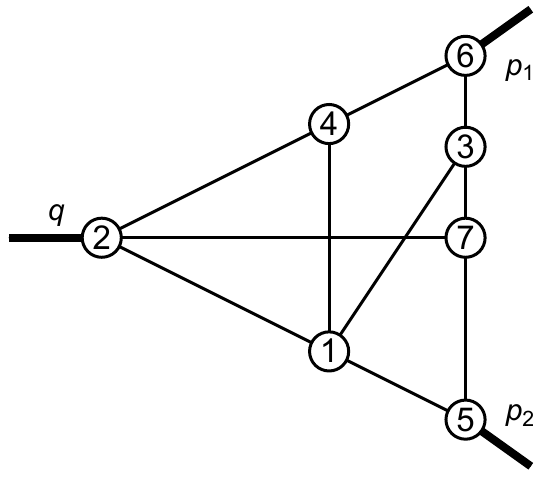}}} \\
        N^{(4)}_{22}&=\vcenter{\hbox{\includegraphics[scale=0.3]{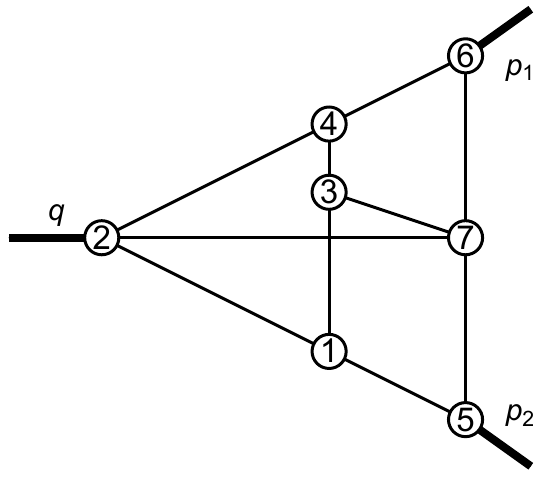}}}, \,
        N^{(4)}_{23}=\vcenter{\hbox{\includegraphics[scale=0.3]{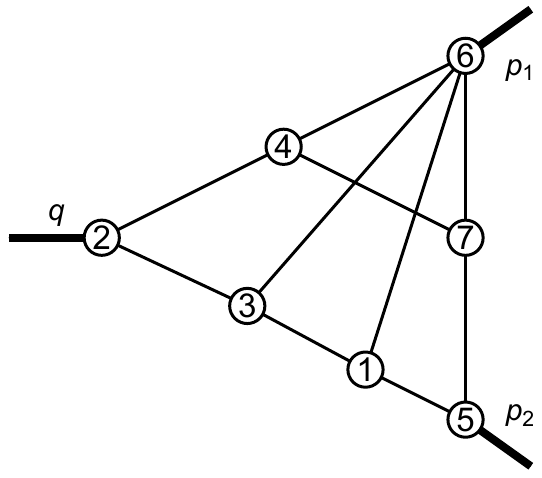}}}
    \end{aligned}
\end{equation}

$N^{(4)}_{23}$ belongs to the special non-planar family and we can easily derive that it equals to the square of two-loop ladders, $F_{2}^{2}$. This is checked by integrating $N^{(4)}_{23}$ in momentum space by \texttt{HyperInt} and its position-space counterpart by \texttt{HyperlogProcedures} respectively. If we further consider the small mass limit of this integral with $q^2=-1$, that is, taking the off-shell legs back to on-shell,
\begin{equation}\label{eq:samemass}
    z=-r,\quad \bar{z}=1+r, \quad\text{  with  }r\to 0,
\end{equation}
the asymptotic behavior of $N^{(4)}_{23}(z,\bar{z})$ is then
\begin{equation}
    N^{(4)}_{23}(z,\bar{z}) = \frac{\pi^{4}}{144}\log^{4}r-\frac{\pi^2}{2}\zeta_{3}\log^{3}r+\left(\frac{\pi^{6}}{90}+9\zeta_{3}^{2}\right)\log^{2}r-\frac{2\pi^{4}}{5}\zeta_{3}\log r+\frac{\pi^8}{225}+\mathcal{O}(r). 
\end{equation}
The coefficient of leading order divergence, $\frac{\pi^{4}}{144}=\frac{5\zeta_{4}}{8}$, agrees with the result calculated from dimensional regularization~\cite{Boels:2017ftb}, which is expected~\cite{Eden:2012fe}.

$N^{(4)}_{22}$ can be calculated from its position-space counterpart by \texttt{HyperlogProcedures}. It equals to $2\times 2$ fishnet, which is the determinant of ladders~\cite{Basso:2017jwq,Basso:2021omx,Aprile:2023gnh}, $F_{1}F_{3}-1/3F_{2}^{2}$\footnote{We note that the arguments of $F_{L}$ here is $z/(z-1)$ and $\bar{z}/(\bar{z}-1)$ instead of $z,\bar{z}$ in eq.~\eqref{eq:FL}, that is, a M\"obius transformation is needed. }. It shows a more complex structure than $N^{(4)}_{22}$ and somehow explains the missing of function basis $F_{1}F_{3}$ in the results. However, for the direct integration of $N^{(4)}_{22}$ in momentum space, \texttt{HyperInt} fails to find a linear reduction order. Nevertheless, we have checked the analytic expression with numerical results evaluated by \texttt{feyntrop}~\cite{Borinsky:2023jdv} in momentum space. They agree well with each other. The leading order coefficient under the limit eq.~\eqref{eq:samemass} also agrees with the result in dimensional regularization. 

The calculation of $N^{(4)}_{20}$ is similar to $N^{(4)}_{22}$. Its position-space counterpart can be calculated by \texttt{HyperlogProcedures}. However, it now involves more complicated multiple polylogarithm with letter $z-\bar{z}$, like $G_{1,\bar{z},\bar{z},\bar{z},0,0,1,1}(z)$ where $G$ is multiple polylogarithm function~\cite{Goncharov:1998kja,Goncharov:2001iea} defined recursively as
\begin{equation}
    G_{a_1,\ldots,a_{n}}(z)=\int_{0}^{z}\frac{\mathrm{d}t}{t-a_{1}}G_{a_2,\ldots,a_{n}}(t).
\end{equation}
Again, the analytic expression agrees well with numerical results evaluated by \texttt{feyntrop} in momentum space. We provide the whole result expressed with $G$ functions of $N_{20}^{(4)}$ in the ancillary file \texttt{N20exp.m}. 

Finally, $N^{(4)}_{19}$ and $N^{(4)}_{21}$ are the two most complicated integrals. Neither of them nor their position space counterparts can be calculated analytically by the current tools. Furthermore, a naive leading singularity analysis of the position-space counterpart of $N^{(4)}_{19}$ shows that $N^{(4)}_{19}$ may involve elliptic integrals. The position-space counterpart of $N^{(4)}_{19}$ is
\begin{equation}
    N^{(4)}_{19}=\int\frac{\mathrm{d}^{4}x_{1}\mathrm{d}^{4}x_{3}\mathrm{d}^{4}x_{4}\mathrm{d}^{4}x_{7}}{(\pi^{2})^{4}}\frac{{\color{orange}x_{28}^{4}x_{56}^{4}}}{x_{12}^{2}x_{13}^{2}x_{14}^{2}x_{15}^{2}x_{24}^{2}x_{36}^{2}x_{37}^{2}x_{46}^{2}x_{47}^{2}x_{57}^{2}{\color{orange}x_{26}^{2}x_{25}^{2}x_{58}^{2}x_{68}^{2}x_{38}^{2}x_{78}^{2}}},
\end{equation}
where the integrand is completed to a conformal integral by the orange part. $x_{2},x_{5},x_{6},x_{8}$ are four external points and $x_{8}$ is at $\infty$\footnote{Note that terms like $x_{26}^{2}$ can be moved out of the integration, since $x_{2}$ and $x_{6}$ are all external points.}. We first cut all propagators for $x_{1}$ and $x_{7}$, which results in two leading singularities $\lambda_{2345}$ and $\lambda_{3458}$ where $\lambda_{ijkl}=\sqrt{\det(x_{mn}^{2})_{m,n=i,j,k,l}}$. They are both the leading singularity of conformal boxes\footnote{This is equivalent to integrating both $x_{1}$ and $x_{7}$ out as conformal boxes.}, which arise from the Jacobian transformation. The remaining cut integral is
\begin{equation}
    \tilde{N}_{19}=\int\frac{\mathrm{d}^{4}x_{3}\mathrm{d}^{4}x_{4}}{(\pi^{2})^{2}}\frac{{\color{orange}x_{28}^{4}x_{56}^{4}}}{x_{24}^{2}x_{36}^{2}x_{46}^{2}\lambda_{2345}\lambda_{3458}{\color{orange}x_{26}^{2}x_{25}^{2}x_{58}^{2}x_{68}^{2}x_{38}^{2}}}.
\end{equation}
Then we consider the cut of $x_{3}$ and $x_{4}$. Under the following cut of propagators
\begin{equation}
   \text{cut}: x_{36}^{2}=x_{38}^{2}=x_{24}^{2}=x_{46}^{2}=0,
\end{equation}
$\lambda_{2345}$ and $\lambda_{3458}$ will degenerate to
\begin{equation}
    \lambda_{2345}|_{\text{cut}} = x_{25}^{2}x_{34}^{2}-x_{23}^{2}x_{45}^{2}, \quad \lambda_{3458}|_{\text{cut}} = x_{35}^{2}x_{48}^{2}-x_{34}^{2}x_{58}^{2}.
\end{equation}
Then the cut of $x_{36}^{2},x_{38}^{2},\lambda_{2345}|_{\text{cut}},\lambda_{3458}|_{\text{cut}}$ for $x_{3}^{\mu}$ results in another Jacobian factor which is,
\begin{equation}
    \begin{aligned}
        \lambda_{68\{245\}\{458\}}|_{\text{cut}}=\sqrt{{\color{lightblue}x_{48}^{2}}\left(x_{28}^{2}{\color{lightblue}x_{45}^{2}}-x_{25}^{2}{\color{lightblue}x_{48}^{2}}\right)\left(x_{28}^{2}x_{56}^{4}{\color{lightblue}x_{45}^{2}}{\color{lightblue}x_{48}^{2}}-x_{25}^{2}x_{56}^{4}{\color{lightblue}x_{48}^{4}}-4x_{26}^{2}x_{58}^{2}x_{68}^{2}{\color{lightblue}x_{45}^{4}}\right)}.
    \end{aligned}
\end{equation}
The corresponding maximal cut integral is
\begin{equation}
    \begin{aligned}
        \tilde{N}_{19}^{\text{max cut}}\propto \int\frac{\mathrm{d}{\color{lightblue}x_{45}^{2}}\mathrm{d}{\color{lightblue}x_{48}^{2}}}{\lambda_{2568}\sqrt{{\color{lightblue}x_{48}^{2}}\left(x_{28}^{2}{\color{lightblue}x_{45}^{2}}-x_{25}^{2}{\color{lightblue}x_{48}^{2}}\right)\left(x_{28}^{2}x_{56}^{4}{\color{lightblue}x_{45}^{2}}{\color{lightblue}x_{48}^{2}}-x_{25}^{2}x_{56}^{4}{\color{lightblue}x_{48}^{4}}-4x_{26}^{2}x_{58}^{2}x_{68}^{2}{\color{lightblue}x_{45}^{4}}\right)}}.
    \end{aligned}
\end{equation}
The polynomial under square root is homogeneous in ${\color{lightblue}x_{45}^{2}}$ and ${\color{lightblue}x_{48}^{2}}$. We can perform a variable transformation to $r\equiv x^{2}_{45}/x_{48}^{2}$ and $x_{48}^{2}$, then cut $x_{48}^{2}$. The polynomial becomes cubic in the remaining integration variable $r$. This signals a elliptic cut of the integral. Performing a similar analysis for $N_{21}^{(4)}$, no elliptic cut has been found. Instead, we find the following leading singularity under the maximal cut of all variables:
\begin{equation}\label{eq:ls}
    \frac{1}{(z-\bar{z})\sqrt{1+z^{2}+\bar{z}^2-2z-2\bar{z}-2z\bar{z}}},\, \frac{x_{26}^{2}x_{58}^{2}}{x_{28}^{2}x_{56}^{2}}\equiv \frac{z\bar{z}}{(1-z)(1-\bar{z})},\, \frac{x_{25}^{2}x_{68}^{2}}{x_{28}^{2}x_{56}^{2}}\equiv \frac{1}{(1-z)(1-\bar{z})}.
\end{equation}
This leading singularity involves a square root of polynomial in $z$ and $\bar{z}$, which shows the complexity of this integral. Nevertheless, we expect this integral to be multiple polylogarithm function after the following variable transformation~\cite{Dlapa:2021qsl,Besier:2018jen}:
\begin{equation}
    z\to -\bar{y}(1+y), \quad \bar{z}\to -y(1+\bar{y}).
\end{equation}
The leading singularity in eq.~\eqref{eq:ls} simplifies to
\begin{equation}
    \frac{1}{y-\bar{y}}
\end{equation}
after considering the Jacobian factor from this variable transformation.

\section{Relation to planar duality by Feynman parameterization}\label{app:proofbyfp}
In Example~\ref{ex:threeloop}, we show the close connection between the generalized Fourier identity and planar duality in the literature~\cite{Golz:2015rea,Borinsky:2021gkd,Borinsky:2022lds}. In this section, we discuss their relations in greater depth.

Generally speaking, the generalized Fourier identity developed in this work relates momentum-space integrals to their position-space counterparts (graphical functions) regardless of the possible underlying topology. It agrees with planar duality only when the momentum-space integrals under consideration admit planar duals. $N_{20}^{(4)}$ in section~\ref{sec:offshellff} is an example which does not admit planar duals, in which case these two concepts are different. In what follows, we restrict ourselves to cases when the planar duals {\it can} exist. 

When the momentum-space integrals are planar graphs, these two are equivalent by definition. Since if a graphical function (which is finite in four dimensions) admits a planar dual, the we can draw its planar dual, which represents a dual integral in momentum space. Then we can use Proposition~\ref{prop:id} to identify this dual integral with its position-space counterpart which shares the same graph. Then the original graphical function is related to the new one which is dual to it in the graph level. The reverse argument from planar duality to generalized Fourier identity is straightforward in the same way.

The more nontrivial case is when the planar dual is not straightfoward like the example in Figure~\ref{fig:example}, Figure~\ref{fig:3lnp} and $N_{23}^{(4)}$ in eq.~\eqref{eq:fourloopnonplanar}. It is not obvious what the planar dual of these integrals would be and how it is related to its planar dual, since the integrals in the momentum space are non-planar, due to the external legs. Now we show how such cases can also be discussed using language of planar duality for momentum-space integrals, taking integrals in Figure~\ref{fig:example} as an example. The key idea is to join the external legs and study the resulting graph. For this example, there {\it exists} such a way that after joining two external legs with the last one, the resulting graph $G$ admits a planar embedding.

We first write down the the Feynman parameterization for the momentum-space integral in Figure~\ref{fig:example},
\begin{equation}
    \begin{aligned}
        \vcenter{\hbox{\includegraphics[scale=0.35]{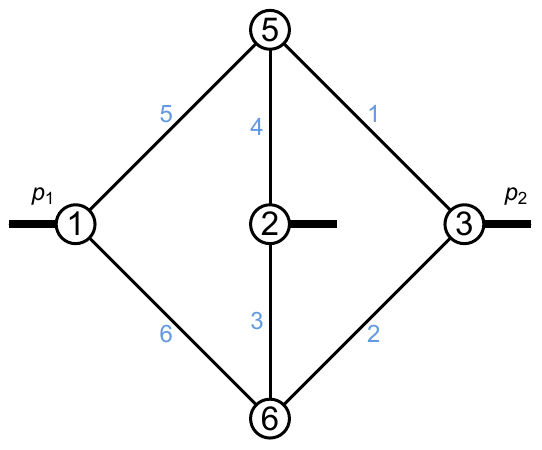}}}=N^{(2)}_{1}=\int[\mathrm{d}\alpha]\frac{1}{Q^2}.
    \end{aligned}
\end{equation}
where $[\mathrm{d}\alpha]$ is a shorthand for the integration measure for projective space of $\alpha_{i}$ and
\begin{equation}
    \begin{aligned}
        Q=&p_{1}^{2}\left(\alpha_{5}\alpha_{6}\left(\alpha_{1}\!+\!\alpha_{2}\!+\!\alpha_{3}\!+\!\alpha_{4}\right)+\alpha_{1}\alpha_{4}\alpha_{6}+\alpha_{2}\alpha_{3}\alpha_{5}\right)+p_{2}^{2}\left(\alpha_{1}\alpha_{2}\left(\alpha_{3}\!+\!\alpha_{4}\!+\!\alpha_{5}\!+\!\alpha_{6}\right)\right.\\
        &\left.+\alpha_{1}\alpha_{3}\alpha_{6}+\alpha_{2}\alpha_{4}\alpha_{5}\right) +q^{2}\left(\alpha_{3}\alpha_{4}\left(\alpha_{1}+\alpha_{2}+\alpha_{5}+\alpha_{6}\right)+\alpha_{2}\alpha_{4}\alpha_{6}+\alpha_{1}\alpha_{3}\alpha_{5}\right).
    \end{aligned}
\end{equation}
Performing Cremona transformation $\alpha\to1/\alpha$, we will arrive at a new representation for the same integral
\begin{equation}
\begin{aligned}
        N^{(2)}_{1}&=\int[\mathrm{d}\alpha]\frac{1}{\tilde{Q}^2}. \\
        \tilde{Q}&=p_{1}^{2}\left(\alpha_{1}\alpha_{4}\left(\alpha_{2}+\alpha_{3}+\alpha_{6}\right)+\alpha_{2}\alpha_{3}\left(\alpha_{1}+\alpha_{4}+\alpha_{5}\right)\right)+p_{2}^{2}\left(\alpha_{3}\alpha_{6}\left(\alpha_{1}+\alpha_{4}+\alpha_{5}\right)\right.\\
        &\left.+\alpha_{4}\alpha_{5}\left(\alpha_{2}+\alpha_{3}+\alpha_{6}\right)\right)+q^{2}\left(\alpha_{2}\alpha_{6}\left(\alpha_{1}+\alpha_{4}+\alpha_{5}\right)+\alpha_{1}\alpha_{5}\left(\alpha_{2}+\alpha_{3}+\alpha_{6}\right)\right).
\end{aligned}
\end{equation}
It corresponds to the following integrals in momentum space
\begin{equation}
    \int[\mathrm{d}\alpha]\frac{1}{\tilde{Q}^2}=\vcenter{\hbox{\includegraphics[scale=0.3]{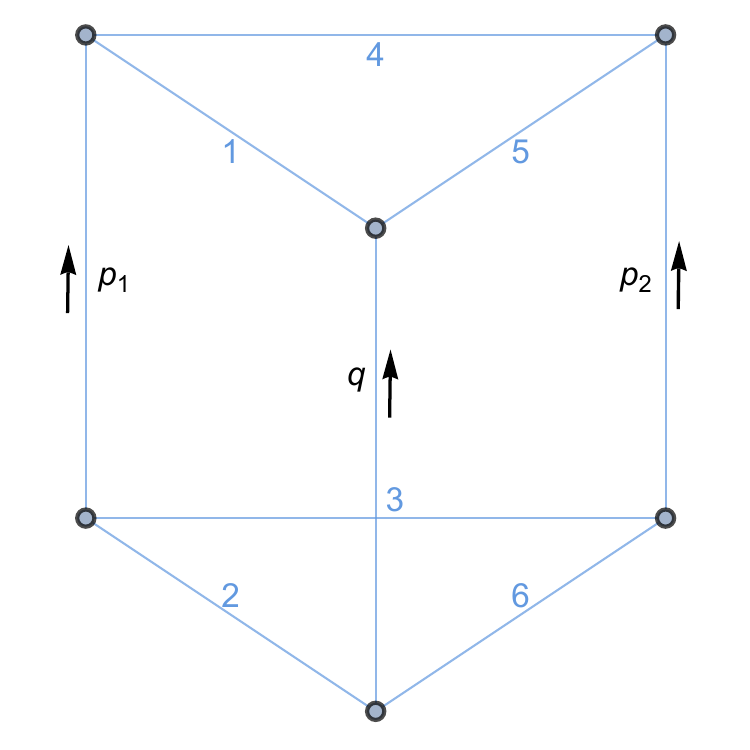}}}
\end{equation}
This is actually the dual graph of $N^{(2)}_{1}$ by first joining three external legs and drawing its dual as follows
\begin{equation}
    \vcenter{\hbox{\includegraphics[scale=0.5]{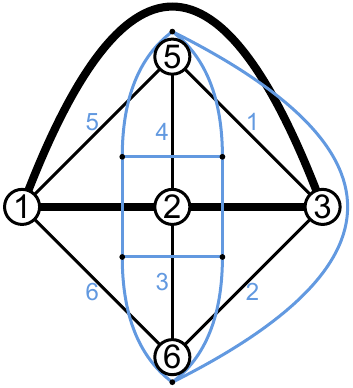}}}
\end{equation}
This is the planar duality directly for the integrals in momentum space. It serves as another way to understand the factorization of non-planar integrals. We do not aim to explain all such identities by using planar duality, but only exploit this example to show that such identities may also be understood by planar duality for the momentum-space integrals which can be proved in Feynman parametric space.

\section{Conclusion and Outlook}\label{sec:conclusion}
In this work, we have proved generalized Fourier identities between massless momentum-space integrals with three off-shell legs and graphical functions (conformal integrals) in position space. Such identities implies the planar duality when the momentum-space integrals admit planar duals and can be generated systematically by graph rules. Especially, We obtained the factorization of a particular family of non-planar integrals, which contributes to the off-shell Sudakov form factors. 

This work is part of the broader effort to uncover the hidden relations among conformal integrals, thereby facilitating their calculation. It is interesting to further explore the implications of such identities. For example, non-planar integral like $N_{23}^{(4)}$ encodes the product of two ladder integrals. Then a natural question is, can the triple products or higher power of ladders be encoded in one single momentum space integral? Since we know the $m\times n$ ($m\le n$) fishnet in momentum space can generate determinant of ladders with $m$ powers. Another related question is how to understand the fact that $N_{22}^{(4)}$ equals to the $2\times 2$ fishnet. Such an equivalence between non-planar integrals and planar integrals is interesting. Is this a coincidence, or is this a consequence of some deep connections between their graph structures which can be exploited to construct similar identities systematically?

Another possible direction is to extend the identities to more general cases. For example, one may ask how to extend the identities to integrals with deformed edge weights when the condition $\nu_1+\nu_2=d/2$ is no long imposed or whether analogous identities exist for higher-point conformal integrals. Finally, in this work, we have focused on conformal integrals, it would also be interesting to investigate whether these identities can be extended to more general integrals without this symmetry. It would be of great interest if more relations are discovered and understood.

\acknowledgments

It is the pleasure of the author to thank Song He, Qinglin Yang and Oliver Schnetz for inspiring discussions and their valuable advice on the structure and presentation of the article. The author is especially grateful to Song He for his encouragement and support in preparing this short note. This work is supported by National Natural Science Foundation of China under Grant No. 12225510.


\bibliographystyle{JHEP}
\bibliography{inspire.bib}

@article{Schnetz:2013hqa,
    author = "Schnetz, Oliver",
    title = "{Graphical functions and single-valued multiple polylogarithms}",
    eprint = "1302.6445",
    archivePrefix = "arXiv",
    primaryClass = "math.NT",
    doi = "10.4310/CNTP.2014.v8.n4.a1",
    journal = "Commun. Num. Theor. Phys.",
    volume = "08",
    pages = "589--675",
    year = "2014"
}

@article{Borinsky:2021gkd,
    author = "Borinsky, Michael and Schnetz, Oliver",
    title = "{Graphical functions in even dimensions}",
    eprint = "2105.05015",
    archivePrefix = "arXiv",
    primaryClass = "hep-th",
    reportNumber = "Nikhef-2021-007",
    doi = "10.4310/CNTP.2022.v16.n3.a3",
    journal = "Commun. Num. Theor. Phys.",
    volume = "16",
    number = "3",
    pages = "515--614",
    year = "2022"
}

@article{Schnetz:2021ebf,
    author = "Schnetz, Oliver",
    title = "{Generalized single-valued hyperlogarithms}",
    eprint = "2111.11246",
    archivePrefix = "arXiv",
    primaryClass = "math-ph",
    month = "11",
    year = "2021"
}

@article{Belitsky:2022itf,
    author = "Belitsky, A. V. and Bork, L. V. and Pikelner, A. F. and Smirnov, V. A.",
    title = "{Exact Off Shell Sudakov Form Factor in N=4 Supersymmetric Yang-Mills Theory}",
    eprint = "2209.09263",
    archivePrefix = "arXiv",
    primaryClass = "hep-th",
    doi = "10.1103/PhysRevLett.130.091605",
    journal = "Phys. Rev. Lett.",
    volume = "130",
    number = "9",
    pages = "091605",
    year = "2023"
}

@article{Belitsky:2023ssv,
    author = "Belitsky, A. V. and Bork, L. V. and Smirnov, V. A.",
    title = "{Off-shell form factor in $ \mathcal{N} $=4 sYM at three loops}",
    eprint = "2306.16859",
    archivePrefix = "arXiv",
    primaryClass = "hep-th",
    doi = "10.1007/JHEP11(2023)111",
    journal = "JHEP",
    volume = "11",
    pages = "111",
    year = "2023"
}

@article{Schnetz:2008mp,
    author = "Schnetz, Oliver",
    title = "{Quantum periods: A Census of phi**4-transcendentals}",
    eprint = "0801.2856",
    archivePrefix = "arXiv",
    primaryClass = "hep-th",
    reportNumber = "FAU-TP3-07-9",
    doi = "10.4310/CNTP.2010.v4.n1.a1",
    journal = "Commun. Num. Theor. Phys.",
    volume = "4",
    pages = "1--48",
    year = "2010"
}

@article{He:2025lzd,
    author = "He, Song and Jiang, Xuhang",
    title = "{Solving Infinite Families of Dual Conformal Integrals and Periods}",
    eprint = "2506.20095",
    archivePrefix = "arXiv",
    primaryClass = "hep-th",
    month = "6",
    year = "2025"
}

@article{Baikov:2010hf,
    author = "Baikov, P. A. and Chetyrkin, K. G.",
    title = "{Four Loop Massless Propagators: An Algebraic Evaluation of All Master Integrals}",
    eprint = "1004.1153",
    archivePrefix = "arXiv",
    primaryClass = "hep-ph",
    reportNumber = "TTP10-18, SFB-CPP-10-24",
    doi = "10.1016/j.nuclphysb.2010.05.004",
    journal = "Nucl. Phys. B",
    volume = "837",
    pages = "186--220",
    year = "2010"
}

@article{Chetyrkin:1980pr,
    author = "Chetyrkin, K. G. and Kataev, A. L. and Tkachov, F. V.",
    title = "{New Approach to Evaluation of Multiloop Feynman Integrals: The Gegenbauer Polynomial x Space Technique}",
    doi = "10.1016/0550-3213(80)90289-8",
    journal = "Nucl. Phys. B",
    volume = "174",
    pages = "345--377",
    year = "1980"
}

@article{Usyukina:1992jd,
    author = "Usyukina, N. I. and Davydychev, Andrei I.",
    title = "{An Approach to the evaluation of three and four point ladder diagrams}",
    reportNumber = "NPI-MSU-92-38-287",
    doi = "10.1016/0370-2693(93)91834-A",
    journal = "Phys. Lett. B",
    volume = "298",
    pages = "363--370",
    year = "1993"
}

@article{Loebbert:2019vcj,
    author = {Loebbert, Florian and M{\"u}ller, Dennis and M{\"u}nkler, Hagen},
    title = "{Yangian Bootstrap for Conformal Feynman Integrals}",
    eprint = "1912.05561",
    archivePrefix = "arXiv",
    primaryClass = "hep-th",
    reportNumber = "HU-EP-19/39",
    doi = "10.1103/PhysRevD.101.066006",
    journal = "Phys. Rev. D",
    volume = "101",
    number = "6",
    pages = "066006",
    year = "2020"
}

@article{Vasiliev:1981dg,
    author = "Vasiliev, A. N. and Pismak, Yu. M. and Khonkonen, Yu. R.",
    title = "{1/$N$ Expansion: Calculation of the Exponents $\eta$ and Nu in the Order 1/$N^2$ for Arbitrary Number of Dimensions}",
    doi = "10.1007/BF01019296",
    journal = "Theor. Math. Phys.",
    volume = "47",
    pages = "465--475",
    year = "1981"
}

@article{Kazakov:1983dyk,
    author = "Kazakov, D. I.",
    title = "{THE METHOD OF UNIQUENESS, A NEW POWERFUL TECHNIQUE FOR MULTILOOP CALCULATIONS}",
    doi = "10.1016/0370-2693(83)90816-X",
    journal = "Phys. Lett. B",
    volume = "133",
    pages = "406--410",
    year = "1983"
}

@article{Isaev:2007uy,
    author = "Isaev, A. P.",
    title = "{Operator approach to analytical evaluation of Feynman diagrams}",
    eprint = "0709.0419",
    archivePrefix = "arXiv",
    primaryClass = "hep-th",
    doi = "10.1134/S1063778808050219",
    journal = "Phys. Atom. Nucl.",
    volume = "71",
    pages = "914--924",
    year = "2008"
}

@article{Chicherin:2012yn,
    author = "Chicherin, D. and Derkachov, S. and Isaev, A. P.",
    title = "{Conformal group: R-matrix and star-triangle relation}",
    eprint = "1206.4150",
    archivePrefix = "arXiv",
    primaryClass = "math-ph",
    doi = "10.1007/JHEP04(2013)020",
    journal = "JHEP",
    volume = "04",
    pages = "020",
    year = "2013"
}

@article{Eden:2011we,
    author = "Eden, Burkhard and Heslop, Paul and Korchemsky, Gregory P. and Sokatchev, Emery",
    title = "{Hidden symmetry of four-point correlation functions and amplitudes in N=4 SYM}",
    eprint = "1108.3557",
    archivePrefix = "arXiv",
    primaryClass = "hep-th",
    reportNumber = "CERN-PH-TH-2011-208, DCPT-11-33, IPHT-T11-91, LAPTH-030-11",
    doi = "10.1016/j.nuclphysb.2012.04.007",
    journal = "Nucl. Phys. B",
    volume = "862",
    pages = "193--231",
    year = "2012"
}

@article{Eden:2012tu,
    author = "Eden, Burkhard and Heslop, Paul and Korchemsky, Gregory P. and Sokatchev, Emery",
    title = "{Constructing the correlation function of four stress-tensor multiplets and the four-particle amplitude in N=4 SYM}",
    eprint = "1201.5329",
    archivePrefix = "arXiv",
    primaryClass = "hep-th",
    reportNumber = "CERN-PH-TH-2012-014, DCPT-12-03, HU-EP-12-03, HU-MATH-2012-27, LAPTH-005-12, IPHT-T12-005",
    doi = "10.1016/j.nuclphysb.2012.04.013",
    journal = "Nucl. Phys. B",
    volume = "862",
    pages = "450--503",
    year = "2012"
}

@article{Bourjaily:2011hi,
    author = "Bourjaily, Jacob L. and DiRe, Alexander and Shaikh, Amin and Spradlin, Marcus and Volovich, Anastasia",
    title = "{The Soft-Collinear Bootstrap: N=4 Yang-Mills Amplitudes at Six and Seven Loops}",
    eprint = "1112.6432",
    archivePrefix = "arXiv",
    primaryClass = "hep-th",
    doi = "10.1007/JHEP03(2012)032",
    journal = "JHEP",
    volume = "03",
    pages = "032",
    year = "2012"
}

@article{Bourjaily:2015bpz,
    author = "Bourjaily, Jacob L. and Heslop, Paul and Tran, Vuong-Viet",
    title = "{Perturbation Theory at Eight Loops: Novel Structures and the Breakdown of Manifest Conformality in N=4 Supersymmetric Yang-Mills Theory}",
    eprint = "1512.07912",
    archivePrefix = "arXiv",
    primaryClass = "hep-th",
    reportNumber = "DCPT-15-75",
    doi = "10.1103/PhysRevLett.116.191602",
    journal = "Phys. Rev. Lett.",
    volume = "116",
    number = "19",
    pages = "191602",
    year = "2016"
}

@article{Bourjaily:2016evz,
    author = "Bourjaily, Jacob L. and Heslop, Paul and Tran, Vuong-Viet",
    title = "{Amplitudes and Correlators to Ten Loops Using Simple, Graphical Bootstraps}",
    eprint = "1609.00007",
    archivePrefix = "arXiv",
    primaryClass = "hep-th",
    reportNumber = "DCPT-16-31",
    doi = "10.1007/JHEP11(2016)125",
    journal = "JHEP",
    volume = "11",
    pages = "125",
    year = "2016"
}

@article{He:2024cej,
    author = "He, Song and Shi, Canxin and Tang, Yichao and Zhang, Yao-Qi",
    title = "{The cusp limit of correlators and a new graphical bootstrap for correlators/amplitudes to eleven loops}",
    eprint = "2410.09859",
    archivePrefix = "arXiv",
    primaryClass = "hep-th",
    doi = "10.1007/JHEP03(2025)192",
    journal = "JHEP",
    volume = "03",
    pages = "192",
    year = "2025"
}

@article{Bourjaily:2025iad,
    author = "Bourjaily, Jacob L. and He, Song and Shi, Canxin and Tang, Yichao",
    title = "{The Four-Point Correlator of Planar sYM at Twelve Loops}",
    eprint = "2503.15593",
    archivePrefix = "arXiv",
    primaryClass = "hep-th",
    month = "3",
    year = "2025"
}

@article{Panzer:2014caa,
    author = "Panzer, Erik",
    title = "{Algorithms for the symbolic integration of hyperlogarithms with applications to Feynman integrals}",
    eprint = "1403.3385",
    archivePrefix = "arXiv",
    primaryClass = "hep-th",
    doi = "10.1016/j.cpc.2014.10.019",
    journal = "Comput. Phys. Commun.",
    volume = "188",
    pages = "148--166",
    year = "2015"
}

@misc{hyperlogprocedures,
  author = {Schnetz, Oliver},
  howpublished = {https://www.math.fau.de/person/oliver-schnetz/},
  title = {HyperlogProcedures},
  year = {2023}
}

@article{Golz:2015rea,
    author = "Golz, Marcel and Panzer, Erik and Schnetz, Oliver",
    title = "{Graphical functions in parametric space}",
    eprint = "1509.07296",
    archivePrefix = "arXiv",
    primaryClass = "math-ph",
    doi = "10.1007/s11005-016-0935-6",
    journal = "Lett. Math. Phys.",
    volume = "107",
    number = "6",
    pages = "1177--1192",
    year = "2017"
}

@article{Borinsky:2022lds,
    author = "Borinsky, Michael and Schnetz, Oliver",
    title = "{Recursive computation of Feynman periods}",
    eprint = "2206.10460",
    archivePrefix = "arXiv",
    primaryClass = "hep-th",
    doi = "10.1007/JHEP08(2022)291",
    journal = "JHEP",
    volume = "22",
    pages = "291",
    year = "2022"
}

@article{Usyukina:1993ch,
    author = "Usyukina, N. I. and Davydychev, Andrei I.",
    title = "{Exact results for three and four point ladder diagrams with an arbitrary number of rungs}",
    reportNumber = "INLO-PUB-1-93",
    doi = "10.1016/0370-2693(93)91118-7",
    journal = "Phys. Lett. B",
    volume = "305",
    pages = "136--143",
    year = "1993"
}

@article{Broadhurst:1993ib,
    author = "Broadhurst, David J.",
    title = "{Summation of an infinite series of ladder diagrams}",
    reportNumber = "OUT-4102-44",
    doi = "10.1016/0370-2693(93)90202-S",
    journal = "Phys. Lett. B",
    volume = "307",
    pages = "132--139",
    year = "1993"
}

@article{Broadhurst:2010ds,
    author = "Broadhurst, David J. and Davydychev, Andrei I.",
    editor = {Bl{\"u}mlein, Johannes and Moch, Sven-Olaf and Riemann, Tord},
    title = "{Exponential suppression with four legs and an infinity of loops}",
    eprint = "1007.0237",
    archivePrefix = "arXiv",
    primaryClass = "hep-th",
    doi = "10.1016/j.nuclphysbps.2010.09.014",
    journal = "Nucl. Phys. B Proc. Suppl.",
    volume = "205-206",
    pages = "326--330",
    year = "2010"
}

@article{Boels:2017ftb,
    author = "Boels, Rutger H. and Huber, Tobias and Yang, Gang",
    title = "{The Sudakov form factor at four loops in maximal super Yang-Mills theory}",
    eprint = "1711.08449",
    archivePrefix = "arXiv",
    primaryClass = "hep-th",
    reportNumber = "SI-HEP-2017-26",
    doi = "10.1007/JHEP01(2018)153",
    journal = "JHEP",
    volume = "01",
    pages = "153",
    year = "2018"
}

@article{Eden:2012fe,
    author = "Eden, Burkhard and Heslop, Paul and Korchemsky, Gregory P. and Smirnov, Vladimir A. and Sokatchev, Emery",
    title = "{Five-loop Konishi in N=4 SYM}",
    eprint = "1202.5733",
    archivePrefix = "arXiv",
    primaryClass = "hep-th",
    reportNumber = "CERN-PH-TH-2012-053, DCPT-12-11, HU-EP-12-07, HU-MATH-2012-04, LAPTH-011-12, IPHT-T12-014",
    doi = "10.1016/j.nuclphysb.2012.04.015",
    journal = "Nucl. Phys. B",
    volume = "862",
    pages = "123--166",
    year = "2012"
}

@article{Basso:2017jwq,
    author = "Basso, Benjamin and Dixon, Lance J.",
    title = "{Gluing Ladder Feynman Diagrams into Fishnets}",
    eprint = "1705.03545",
    archivePrefix = "arXiv",
    primaryClass = "hep-th",
    reportNumber = "SLAC-PUB-16967",
    doi = "10.1103/PhysRevLett.119.071601",
    journal = "Phys. Rev. Lett.",
    volume = "119",
    number = "7",
    pages = "071601",
    year = "2017"
}

@article{Basso:2021omx,
    author = "Basso, Benjamin and Dixon, Lance J. and Kosower, David A. and Krajenbrink, Alexandre and Zhong, De-liang",
    title = "{Fishnet four-point integrals: integrable representations and thermodynamic limits}",
    eprint = "2105.10514",
    archivePrefix = "arXiv",
    primaryClass = "hep-th",
    reportNumber = "SLAC{\textendash}PUB{\textendash}17600, SLAC--PUB--17600",
    doi = "10.1007/JHEP07(2021)168",
    journal = "JHEP",
    volume = "07",
    pages = "168",
    year = "2021"
}

@article{Aprile:2023gnh,
    author = "Aprile, Francesco and Olivucci, Enrico",
    title = "{Multipoint fishnet Feynman diagrams: Sequential splitting}",
    eprint = "2307.12984",
    archivePrefix = "arXiv",
    primaryClass = "hep-th",
    doi = "10.1103/PhysRevD.108.L121902",
    journal = "Phys. Rev. D",
    volume = "108",
    number = "12",
    pages = "L121902",
    year = "2023"
}

@article{Borinsky:2023jdv,
    author = "Borinsky, Michael and Munch, Henrik J. and Tellander, Felix",
    title = "{Tropical Feynman integration in the Minkowski regime}",
    eprint = "2302.08955",
    archivePrefix = "arXiv",
    primaryClass = "hep-ph",
    reportNumber = "DESY-23-026",
    doi = "10.1016/j.cpc.2023.108874",
    journal = "Comput. Phys. Commun.",
    volume = "292",
    pages = "108874",
    year = "2023"
}

@article{Chetyrkin:1981qh,
    author = "Chetyrkin, K. G. and Tkachov, F. V.",
    title = "{Integration by parts: The algorithm to calculate $\beta$-functions in 4 loops}",
    doi = "10.1016/0550-3213(81)90199-1",
    journal = "Nucl. Phys. B",
    volume = "192",
    pages = "159--204",
    year = "1981"
}

@article{Tkachov:1981wb,
    author = "Tkachov, F. V.",
    title = "{A theorem on analytical calculability of 4-loop renormalization group functions}",
    doi = "10.1016/0370-2693(81)90288-4",
    journal = "Phys. Lett. B",
    volume = "100",
    pages = "65--68",
    year = "1981"
}

@article{Laporta:2000dsw,
    author = "Laporta, S.",
    title = "{High-precision calculation of multiloop Feynman integrals by difference equations}",
    eprint = "hep-ph/0102033",
    archivePrefix = "arXiv",
    doi = "10.1142/S0217751X00002159",
    journal = "Int. J. Mod. Phys. A",
    volume = "15",
    pages = "5087--5159",
    year = "2000"
}

@article{Laporta:2000dc,
    author = "Laporta, S.",
    title = "{Calculation of master integrals by difference equations}",
    eprint = "hep-ph/0102032",
    archivePrefix = "arXiv",
    doi = "10.1016/S0370-2693(01)00256-8",
    journal = "Phys. Lett. B",
    volume = "504",
    pages = "188--194",
    year = "2001"
}

@article{Larsen:2015ped,
    author = "Larsen, Kasper J. and Zhang, Yang",
    title = "{Integration-by-parts reductions from unitarity cuts and algebraic geometry}",
    eprint = "1511.01071",
    archivePrefix = "arXiv",
    primaryClass = "hep-th",
    doi = "10.1103/PhysRevD.93.041701",
    journal = "Phys. Rev. D",
    volume = "93",
    number = "4",
    pages = "041701",
    year = "2016"
}

@article{Ita:2015tya,
    author = "Ita, Harald",
    title = "{Two-loop Integrand Decomposition into Master Integrals and Surface Terms}",
    eprint = "1510.05626",
    archivePrefix = "arXiv",
    primaryClass = "hep-th",
    reportNumber = "FR-PHENO-2015-011",
    doi = "10.1103/PhysRevD.94.116015",
    journal = "Phys. Rev. D",
    volume = "94",
    number = "11",
    pages = "116015",
    year = "2016"
}

@article{Bohm:2017qme,
    author = {B{\"o}hm, Janko and Georgoudis, Alessandro and Larsen, Kasper J. and Schulze, Mathias and Zhang, Yang},
    title = "{Complete sets of logarithmic vector fields for integration-by-parts identities of Feynman integrals}",
    eprint = "1712.09737",
    archivePrefix = "arXiv",
    primaryClass = "hep-th",
    reportNumber = "MITP-17-104, UUITP-44-17, MITP/17-104, UUITP-44/17",
    doi = "10.1103/PhysRevD.98.025023",
    journal = "Phys. Rev. D",
    volume = "98",
    number = "2",
    pages = "025023",
    year = "2018"
}

@article{Mastrolia:2018uzb,
    author = "Mastrolia, Pierpaolo and Mizera, Sebastian",
    title = "{Feynman Integrals and Intersection Theory}",
    eprint = "1810.03818",
    archivePrefix = "arXiv",
    primaryClass = "hep-th",
    doi = "10.1007/JHEP02(2019)139",
    journal = "JHEP",
    volume = "02",
    pages = "139",
    year = "2019"
}

@article{Frellesvig:2019uqt,
    author = "Frellesvig, Hjalte and Gasparotto, Federico and Mandal, Manoj K. and Mastrolia, Pierpaolo and Mattiazzi, Luca and Mizera, Sebastian",
    title = "{Vector Space of Feynman Integrals and Multivariate Intersection Numbers}",
    eprint = "1907.02000",
    archivePrefix = "arXiv",
    primaryClass = "hep-th",
    doi = "10.1103/PhysRevLett.123.201602",
    journal = "Phys. Rev. Lett.",
    volume = "123",
    number = "20",
    pages = "201602",
    year = "2019"
}

@article{Song:2025pwy,
    author = "Song, Zhuo-Yang and Yang, Tong-Zhi and Cao, Qing-Hong and Luo, Ming-xing and Zhu, Hua Xing",
    title = "{Explainable AI-assisted Optimization for Feynman Integral Reduction}",
    eprint = "2502.09544",
    archivePrefix = "arXiv",
    primaryClass = "hep-ph",
    reportNumber = "ZU-TH 07/25",
    month = "2",
    year = "2025"
}

@article{Zeng:2025xbh,
    author = "Zeng, Mao",
    title = "{Reinforcement Learning and Metaheuristics for Feynman Integral Reduction}",
    eprint = "2504.16045",
    archivePrefix = "arXiv",
    primaryClass = "hep-ph",
    month = "4",
    year = "2025"
}

@article{Drummond:2006rz,
    author = "Drummond, J. M. and Henn, J. and Smirnov, V. A. and Sokatchev, E.",
    title = "{Magic identities for conformal four-point integrals}",
    eprint = "hep-th/0607160",
    archivePrefix = "arXiv",
    reportNumber = "LAPTH-1159-06",
    doi = "10.1088/1126-6708/2007/01/064",
    journal = "JHEP",
    volume = "01",
    pages = "064",
    year = "2007"
}

@article{Caron-Huot:2021usw,
    author = "Caron-Huot, Simon and Coronado, Frank",
    title = "{Ten dimensional symmetry of $ \mathcal{N} $ = 4 SYM correlators}",
    eprint = "2106.03892",
    archivePrefix = "arXiv",
    primaryClass = "hep-th",
    doi = "10.1007/JHEP03(2022)151",
    journal = "JHEP",
    volume = "03",
    pages = "151",
    year = "2022"
}

@article{He:2025vqt,
    author = "He, Song and Jiang, Xuhang and Liu, Jiahao and Zhang, Yao-Qi",
    title = "{Notes on conformal integrals: Coulomb branch amplitudes, magic identities, and bootstrap}",
    eprint = "2502.08871",
    archivePrefix = "arXiv",
    primaryClass = "hep-th",
    doi = "10.1103/gmp7-r9dz",
    journal = "Phys. Rev. D",
    volume = "112",
    number = "7",
    pages = "076012",
    year = "2025"
}

@article{Hu:2018liw,
    author = "Hu, Simone and Schnetz, Oliver and Shaw, Jim and Yeats, Karen",
    title = "{Further investigations into the graph theory of $\phi^4$-periods and the $c_2$ invariant}",
    eprint = "1812.08751",
    archivePrefix = "arXiv",
    primaryClass = "hep-th",
    doi = "10.4171/aihpd/123",
    journal = "Ann. Inst. H. Poincare D Comb. Phys. Interact.",
    volume = "9",
    number = "3",
    pages = "473--524",
    year = "2022"
}

@article{Schnetz:2025mjw,
    author = "Schnetz, Oliver",
    title = "{The five-twist identity for Feynman periods}",
    eprint = "2505.02578",
    archivePrefix = "arXiv",
    primaryClass = "hep-th",
    month = "5",
    year = "2025"
}

@article{Broadhurst:1995km,
    author = "Broadhurst, David J. and Kreimer, D.",
    editor = "Denby, Bruce H. and Perret-Gallix, D.",
    title = "{Knots and numbers in Phi**4 theory to 7 loops and beyond}",
    eprint = "hep-ph/9504352",
    archivePrefix = "arXiv",
    reportNumber = "OUT-4102-57, MZ-TH-95-12A, UTAS-PHYS-95-12",
    doi = "10.1142/S012918319500037X",
    journal = "Int. J. Mod. Phys. C",
    volume = "6",
    pages = "519--524",
    year = "1995"
}

@article{Kreimer:1996js,
    author = "Kreimer, Dirk",
    title = "{Renormalization and knot theory}",
    eprint = "q-alg/9607022",
    archivePrefix = "arXiv",
    reportNumber = "MZ-TH-96-18",
    doi = "10.1142/S0218216597000315",
    journal = "J. Knot Theor. Ramifications",
    volume = "6",
    pages = "479--581",
    year = "1997"
}

@article{Broadhurst:1996az,
    author = "Broadhurst, David J.",
    title = "{On the enumeration of irreducible k fold Euler sums and their roles in knot theory and field theory}",
    eprint = "hep-th/9604128",
    archivePrefix = "arXiv",
    reportNumber = "OUT-4102-62",
    month = "4",
    year = "1996"
}

@article{Broadhurst:1996kc,
    author = "Broadhurst, David J. and Kreimer, D.",
    title = "{Association of multiple zeta values with positive knots via Feynman diagrams up to 9 loops}",
    eprint = "hep-th/9609128",
    archivePrefix = "arXiv",
    reportNumber = "UTAS-PHYS-96-44, OUT-4102-64, MZ-TH-96-25",
    doi = "10.1016/S0370-2693(96)01623-1",
    journal = "Phys. Lett. B",
    volume = "393",
    pages = "403--412",
    year = "1997"
}

@article{Broadhurst:1986bx,
    author = "Broadhurst, David J.",
    title = "{Exploiting the 1.440 Fold Symmetry of the Master Two Loop Diagram}",
    doi = "10.1007/BF01552503",
    journal = "Z. Phys. C",
    volume = "32",
    pages = "249--253",
    year = "1986"
}

@article{Brown:2009ta,
    author = "Brown, Francis C. S.",
    title = "{On the periods of some Feynman integrals}",
    eprint = "0910.0114",
    archivePrefix = "arXiv",
    primaryClass = "math.AG",
    month = "10",
    year = "2009"
}

@article{Brown:2010bw,
    author = "Brown, Francis and Schnetz, Oliver",
    title = "{A K3 in $\phi^4$}",
    eprint = "1006.4064",
    archivePrefix = "arXiv",
    primaryClass = "math.AG",
    doi = "10.1215/00127094-1644201",
    journal = "Duke Math. J.",
    volume = "161",
    number = "10",
    pages = "1817--1862",
    year = "2012"
}

@article{Georgoudis:2021onj,
    author = "Georgoudis, Alessandro and Gon{\c{c}}alves, Vasco and Panzer, Erik and Pereira, Raul and Smirnov, Alexander V. and Smirnov, Vladimir A.",
    title = "{Glue-and-cut at five loops}",
    eprint = "2104.08272",
    archivePrefix = "arXiv",
    primaryClass = "hep-ph",
    doi = "10.1007/JHEP09(2021)098",
    journal = "JHEP",
    volume = "09",
    pages = "098",
    year = "2021"
}

@article{Goncharov:1998kja,
    author = "Goncharov, Alexander B.",
    title = "{Multiple polylogarithms, cyclotomy and modular complexes}",
    eprint = "1105.2076",
    archivePrefix = "arXiv",
    primaryClass = "math.AG",
    doi = "10.4310/MRL.1998.v5.n4.a7",
    journal = "Math. Res. Lett.",
    volume = "5",
    pages = "497--516",
    year = "1998"
}

@article{Goncharov:2001iea,
    author = "Goncharov, A. B.",
    title = "{Multiple polylogarithms and mixed Tate motives}",
    eprint = "math/0103059",
    archivePrefix = "arXiv",
    month = "3",
    year = "2001"
}

@article{Dlapa:2021qsl,
    author = "Dlapa, Christoph and Li, Xiaodi and Zhang, Yang",
    title = "{Leading singularities in Baikov representation and Feynman integrals with uniform transcendental weight}",
    eprint = "2103.04638",
    archivePrefix = "arXiv",
    primaryClass = "hep-th",
    reportNumber = "MPP-2021-23, PCFT-21-10, USTC-ICTS-21-10",
    doi = "10.1007/JHEP07(2021)227",
    journal = "JHEP",
    volume = "07",
    pages = "227",
    year = "2021"
}

@article{Besier:2018jen,
    author = "Besier, Marco and Van Straten, Duco and Weinzierl, Stefan",
    title = "{Rationalizing roots: an algorithmic approach}",
    eprint = "1809.10983",
    archivePrefix = "arXiv",
    primaryClass = "hep-th",
    doi = "10.4310/CNTP.2019.v13.n2.a1",
    journal = "Commun. Num. Theor. Phys.",
    volume = "13",
    pages = "253--297",
    year = "2019"
}

@article{Kardos:2025klp,
    author = "Kardos, Adam and Moch, Sven-Olaf and Schnetz, Oliver",
    title = "{HyperFORM -- a FORM package for parametric integration with hyperlogarithms}",
    eprint = "2511.19992",
    archivePrefix = "arXiv",
    primaryClass = "hep-ph",
    month = "11",
    year = "2025"
}


\end{document}